\documentclass[11pt,a4paper]{article}
\pdfoutput=1

\usepackage{amsfonts}
\usepackage{jheppub}
\usepackage{amsmath}
\usepackage{amssymb}
\usepackage{xcolor}
\usepackage{float}
\usepackage{graphicx}%
\usepackage[normalem]{ulem}
\newcommand{\stkout}[1]{\ifmmode\text{\sout{\ensuremath{#1}}}\else\sout{#1}\fi}
\usepackage{url}	
\usepackage{braket}
\usepackage[mathscr]{eucal}
\usepackage{bbm}	
\usepackage{youngtab}
\usepackage{ytableau}
\usepackage{bm}

\makeatletter
\newenvironment{equations}[1][]{\subequations\ifx\relax#1\relax\else\label{#1}\fi\align\ignorespaces}{\endalign\ignorespacesafterend\endsubequations}
\def\@spliteq#1{\begin{equation}\begin{split}#1\end{split}\end{equation}}
\def\splitequation{\collect@body\@spliteq}

\makeatother

\newcommand{\diff}{\mathrm{d}}
\newcommand{\de}{\partial}
\newcommand{\vepsilon}{\varepsilon}
\newcommand{\vtheta}{\vartheta}

\newcommand{\vrho}{\varrho}

\makeatletter
\@addtoreset{equation}{section}

\makeatother

\newcommand{\nn}{\nonumber}
\newlength{\sswidth}

\title{Teleparallel formulation of $E_{8(8)}$ Exceptional Field Theory}

\author[a]{Davide Rovere}

\affiliation[a]{Centre de Physique Théorique, CNRS, Institut Polytechnique de Paris, 91128 Palaiseau Cedex, France
 \vspace{0.1cm}}

\emailAdd{davide.rovere@polytechnique.edu}

\abstract{The potential of $E_{8(8)} \times \mathbb{R}^+$ exceptional field theory, suited for studying three-dimensional maximally supersymmetric compactifications of higher-dimensional supergravities, is derived in teleparallel formulation. This means that, starting from an ansatz of the most general combination of the quadratic scalar contractions of the intrinsic torsion, which is manifestly invariant under generalised diffeomorphisms, one selects the potential by imposing the invariance under infinitesimal rotation of the maximal compact subgroup of the duality group. The teleparallel formulation makes manifest at geometric level the preeminent r\^ole of the intrinsic torsion in studying compactifications of higher-dimensional supergravities, where the intrinsic torsion is identified with the embedding tensor of the resulting gauged supergravity.}

\begin{document}

\maketitle

\section{Introduction}

The aim of this note is to study explicitly  the teleparallel formulation of the potential of exceptional field theory with $n=8$ internal dimensions \cite{Hohm:2014fxa}, continuing the program started in \cite{Cederwall:2023xbj}.

\emph{Exceptional field theories} (ExFT) are reformulations of 10d or 11d supergravities introduced to make manifest the duality symmetries of their lower-dimensional maximally supersymmetric compactifications {\cite{Hohm:2013vpa, Hohm:2013uia, Hohm:2014fxa}. A duality is a symmetry of the classical equations of motion of two theories, which sends one theory into another while preserving the equations of motion. When 11d supergravity is compactified up to $11-n=5,4,3$ dimensions, the lower-dimensional theories enjoy a duality under the maximal non-compact form of the exceptional Lie groups $E_{n(n)}$, where $n=6,7,8$ is respectively the dimension of the internal space \cite{Julia:1980gr, Cremmer:1997ct}.

In ExFT the fields are repackaged in a covariant way with respect to the exceptional groups \emph{before} the reduction. This is possible by means of  
an extension of generalised geometry \cite{Hitchin:2003cxu,Gualtieri:2003dx, Hitchin:2005in, Gualtieri:2007ng, Hitchin:2010qz}, dubbed ``exceptional geometry" \cite{Hull:2007zu, Grana:2008yw, PiresPacheco:2008qik, Coimbra:2011ky, Coimbra:2011nw, Coimbra:2012af, Strickland-Constable:2013xta}. The characterising feature of generalised geometry is the combination of vector fields and 
$p$-forms into single generalised vector fields: by means of a deformation of the Lie derivative, called \emph{generalised Lie derivative} $L$, the vector component of generalised vectors generates the usual diffeomorphisms, and the $p$-form components generate $p$-form gauge transformations. 

In exceptional geometric formulations of 11d supergravity, one considers generalised vectors with vector fields and two-forms as components, the latter being the gauge parameter of the three-form gauge transformation of 11d supergravity, and, in addition to these, the generators of the dimensional-dependent dual gauge symmetries are also included \cite{Hull:2007zu,PiresPacheco:2008qik, Coimbra:2011ky, Coimbra:2011nw, Coimbra:2012af}. In this way, the generalised vectors transform in the fundamental representation of $E_{n(n)}$, according to its decomposition with respect to $\text{GL}(n)$.

Scherk and Schwarz showed that, if the compactified space is parallelisable (that is, it admits an absolute frame) with \emph{constant} intrinsic torsion, then the solutions of the reduced theory are also solutions of the higher one (\emph{consistent truncation}), and the absolute frame gives the dependence on the internal coordinates in the reduced fields \cite{Scherk:1978ta, Scherk:1979zr}. In particular, this is the case when the internal space is a Lie group, since Lie groups are parallelisable manifolds whose intrinsic torsion is (minus) the structure constants. If $e_a = e^\mu{}_a\,\de_\mu$ denotes the absolute frame, then the components $T_a{}^c{}_b$ of the intrinsic torsion are defined as (minus) the coefficient of commutator $[e_a,e_b]$ (the same as the Lie derivative $\mathcal{L}_{e_a} e_b$) written as a linear combination of the absolute frame itself 
\begin{equation}
[e_a,e_b] = \mathcal{L}_{e_a} e_b = -T_a{}^c{}_b\,e_c.
\end{equation}
The \emph{generalised intrinsic torsion} is  similarly defined by replacing the frame $e_a$ and the Lie derivative $\mathcal{L}$ with their generalised siblings $E_A$ and $L$:
\begin{equation}
L_{E_A}\,E_B = -T_A{}^C{}_B\,E_C,
\end{equation}
and the Scherk-Schwarz reduction condition is extended by requiring that the internal manifold is parallelisable in generalised sense with constant generalised intrinsic torsion. The latter is identified with the components $X_{AB}{}^C$ of the embedding tensor, which completely characterises the gauging of the reduced gauged supergravity  \cite{Samtleben:2008pe, Trigiante:2016mnt}.

For compactifications on internal manifolds other than hypertori, the reduced theory is in general characterised by a \emph{scalar potential}, and a subgroup of the global duality group acts as a gauge group, whose gauge fields are chosen among the Abelian vector fields of the supergravity multiplet.

The \emph{embedding tensor} is a matrix $\vtheta_A{}^\alpha$, where the index $A$ is in the fundamental representation of the duality group and $\alpha$ onto the adjoint one, which is used to project the fundamental index in the adjoint one. Its components are defined as
\begin{equation}
X_{AB}{}^C = \vtheta_A{}^\alpha\,(t_\alpha)^C{}_B,
\end{equation}
where $\{t_\alpha\}$ are the generators of the Lie algebra of the duality group. If the embedding tensor is viewed as a linear map sending elements in the fundamental representation to elements in the adjoint one, its image is required to be the Lie algebra of the gauge group. This gives rise to a closure condition (``quadratic constraint"), which can be written as
\begin{equation}
[X_A,X_B]_E{}^D=-X_{AB}{}^C\,X_{CE}{}^D.
\end{equation}
Moreover, the compatibility with supersymmetry restricts the representations in which the embedding tensor transforms, which are contained in the tensor product of the fundamental and the adjoint one (``linear constraint").

Following this perspective, a geometric interpretation of the embedding tensor of the reduced theory is gained: the embedding tensor is the generalised intrinsic torsion of the internal manifold, when the latter is constant. Namely, the embedding tensor and the generalised intrinsic torsion sit in the same representations of the duality group and the linear constraint is a consequence of the fact that the generalised intrinsic torsion is a covariant tensor with respect to the generalised Lie derivative. Moreover, the quadratic constraint of the embedding tensor has its geometric analogue in the Bianchi identity satisfied by the torsion.

The preeminent r\^ole of the intrinsic torsion in exceptional geometry was emphasised in \cite{Cederwall:2021xqi}, and further developed in \cite{Cederwall:2023xbj, Bossard:2024gry},  quite later than the first introduction of generalised geometry, since in the usual formulation of Einstein gravity one takes a torsionless connection, the Levi-Civita connection, assigning the gravitational degrees of freedom to its curvature, the Riemann tensor $R^\mu{}_{\nu\vrho\sigma}$. But the extension of this formulation in generalised geometry is plagued by a problem, since it is not possible to define a Riemann tensor in generalised geometry \cite{Coimbra:2011ky}. Nevertheless, it is  possible to define a Ricci tensor, and hence an Einstein-Hilbert action. 

There is an alternative formulation of the Einstein-Hilbert action, even in ordinary geometry, in terms of intrinsic torsion. This formulation, known as the \emph{teleparallel equivalent of General Relativity} \cite{Einstein1928, Weitzenbock1928, deAndrade:1997gka, Maluf:2013gaa, Cederwall:2015ica, Rovere:2025nfj}, is based on the interpretation of intrinsic torsion as the torsion of the Maurer-Cartan form of the frame, viewed as a connection, and called, in this context, the \emph{Weitzenb\"ock connection}:
\begin{equation}
W_\mu{}^\vrho{}_\nu = e^\vrho{}_a\,\de_\mu\,e_\nu{}^a.
\end{equation}
While the Levi-Civita connection is torsionless but has curvature, the Weitzenb\"ock connection is flat but has torsion, which is its antisymmetric part:
\begin{equation}
T_\mu{}^\vrho{}_\nu = W_{\mu}{}^\vrho{}_\nu - W_\nu{}^\vrho{}_\mu.
\end{equation}
In teleparallel formulation, the gravitational effects are explained as torsion effects, instead of effects due to spacetime curvature. Since the Riemann tensor is second order in the derivatives of the metric, whereas the torsion is first order in the derivatives of the frame, the analogue of the Hilbert-Einstein action must be quadratic in the torsion. Since the torsion is a rank-three tensor, with two antisymmetrised indices, there are only three independent quadratic scalar contractions, which can be chosen as $T^{\mu\nu\vrho}\,T_{\mu\nu\vrho}$, $T^{\mu\nu\vrho}\,T_{\mu\vrho\nu}$, and $T^{\mu\nu}{}_\nu\,T_{\mu\vrho}{}^\vrho$. Each of them is generally invariant, since the torsion is a covariant tensor, but they are not local Lorentz invariant. Nevertheless, there is a unique combination, up to total derivatives, of the three scalar contractions which is Lorentz-invariant. This combination turns out to be equivalent to the Ricci scalar $\mathscr{R}$, up to a total derivative:
\begin{equation}
\int\diff^d x\,e\,\mathscr{R} = \int\diff^d x\,e\,\bigg{[}-\frac{1}{4}\,T^{\mu\nu\vrho}\,T_{\mu\nu\vrho} - \frac{1}{2}\,T^{\mu\nu\vrho}\,T_{\mu\vrho\nu} + T_{\mu\nu}{}^\nu\,T^{\mu\vrho}{}_\vrho\bigg{]}.
\end{equation}

In \cite{Cederwall:2023xbj} the teleparallel formulation for exceptional geometry is studied up to $n\leqslant 7$ and in \cite{Bossard:2024gry} the infinite-dimensional Kac-Moody extension, which is relevant in compactifications up to 2d and which relies on the infinite-dimensional exceptional group $E_9$, is explored. 

The main novelty of $n=8$ exceptional geometry is the presence of the so-called ``ancillary parameters". This is due to the fact that the smallest non-trivial representation of the algebra of $E_{8(8)}$, which is the adjoint representation, is larger than the number of 11d supergravity degrees of freedom, so that one is forced to include additional parameters in the generalised Lie derivative in order to get a consistent derivative \cite{Strickland-Constable:2013xta, Hohm:2014fxa, Cederwall:2015ica}.

The starting point for the ExFT potential in teleparallel formulation is the most general quadratic ansatz, which reads:
\begin{align}
V = r^{-3} &\,[\alpha\,\vtheta^2 + \beta\,\vtheta'_M\,\vtheta'^M + 
\gamma\,\vtheta''_{MN}\,\vtheta''^{MN} + \vepsilon\,\vtheta'_M\,\vtheta'_N\,M^{MN} \nn\\
+&\, \zeta\,\vtheta''_{MP}\,\vtheta''_{NQ}\,M^{MN}\,M^{PQ} + \xi\,\vtheta''_{MP}\,\vtheta''^M{}_Q\,M^{PQ}],
\end{align}
where $\alpha, \beta, \dots$ are arbitrary constants; $\vtheta$, $\vtheta'_M$, and $\vtheta''_{MN}$ are the components of the intrinsic torsion, transforming respectively in the representations $\bf{1}$, $\bf{248}$, and $\bf{3875}$ of the algebra of $E_{8(8)}$; the symmetric matrix $M_{MN}$, defined in \eqref{Mmatrix}, is the same as in \cite{Hohm:2014fxa}; and the $r$ factor, parametrising the $\mathbb{R}^+$ of the duality group -- the so-called ``trombone duality" \cite{Cremmer:1997xj, LeDiffon:2008sh} -- ensures that the whole expression is a scalar density with respect to generalised diffeomorphisms. Imposing the invariance under the infinitesimal local $\text{SO}(16)$ rotations, where $\text{SO}(16)$ is the maximal 
compact subgroup of $E_{8(8)}$, 
\begin{equation}
\delta_L\,V = 0,
\end{equation}
and using the section constraint, which implies the following quadratic identity, \emph{not} implied by the quadratic constraint
\cite{Eloy:2023zzh, Inverso:2024xok},
\begin{equation}
\vtheta''_{MN}\,\vtheta''^{MN} - 21\,\vtheta'_M\,\vtheta'^M + 280\,\vtheta^2 \approx 0,
\end{equation}
one gets a unique result for the potential:
\begin{align}
V &= r^{-3}[-32\,\vtheta^2 + \tfrac{5}{2}\,\vtheta'_M\,\vtheta'^M -\tfrac{3}{28}\,\vtheta''_{MN}\,\vtheta''^{MN} - 
\tfrac{3}{2}\,\vtheta'_M\,\vtheta'_N\,M^{MN} \nn\\
& + \tfrac{1}{28}\,\vtheta''_{MP}\,\vtheta''_{NQ}\,M^{MN}\,M^{PQ} + \tfrac{1}{2}\,\vtheta''_{MP}\,\vtheta''^M{}_Q\,M^{PQ}],
\end{align}
as it will be shown in this note. The result is equivalent, up to a total derivative, to the potential originally found in \cite{Hohm:2014fxa}, as it will be shown too. Moreover, this potential can be compared with the potential for three-dimensional gauged supergravities, first studied in \cite{Nicolai:2000sc, Nicolai:2001sv}.

The teleparallel formalism for generalised geometry is presented in general in \cite{Cederwall:2021xqi,  Cederwall:2023xbj}. In particular, in \cite{Cederwall:2023xbj} there is also brief section devoted to the case of $E_8$, with information relevant for their analysis, extracted using group theory. In the present work we explicitly work out the dynamics of $E_{8(8)}$ ExFT, obtaining the potential in the teleparallel approach from symmetry principles. As mentioned before, the generalised Lie derivative of the $n<8$ cases must be supplemented by the ancillary transformations in the $n=8$ case to ensure closure. The presence of the ancillary parameter gives rise to an additional term in the torsion, which is absent in the $n<8$ cases. In this way, the ``big torsion" and ``small torsion" used in \cite{ Cederwall:2023xbj} can be assembled into the full generalised torsion $T_{A}{}^C{}_B$ defined above, and the Bianchi identity, which we will explicitly write down and study in its consequences, follows directly from the Leibniz identity of the generalised Lie derivative, just as in ordinary geometry.

The structure of the paper is the following: in Section \ref{Sec1} the teleparallel formulation of General Relativity is reviewed in a pedagogical way, with particular emphasis on the aspects which have to be extended in the generalised case; in Section \ref{Sec2} there is a summary of the features of $n=8$ exceptional geometry relevant for this note (in particular the necessary Bianchi identities are derived); Section \ref{Sec3} contains the full derivation of the potential of $n=8$ exceptional field theory in teleparallel formulation; Section \ref{Sec4} makes a comparison with the metric formulation in which the potential was first presented; in Section \ref{Sec5} there is a comparison with the potential of three-dimensional gauged supergravities; in Section \ref{Sec6} the possibility of weakening the section constraint to eliminate an ambiguity in the potential is discussed; in Appendix \ref{App1} there is other useful material on $n=8$ exceptional geometry; in Appendix \ref{AppDiff} there are details on generalised diffeomorphisms; finally Appendix \ref{App3} is devoted to a detailed derivation of the relevant Bianchi identities.

\section{Teleparallel formulation of ordinary geometry}\label{Sec1}

In this section the teleparallel formulation in ordinary geometry is reviewed \cite{Einstein1928, Weitzenbock1928, deAndrade:1997gka, Maluf:2013gaa, Cederwall:2015ica, Rovere:2025nfj}. Consider a $d$-dimensional manifold $\mathscr{M}$, equipped with a metric structure. Denote by $e_\mu{}^a$ the components of the vielbein, and with $e^\mu{}_a$ the components of the inverse vielbein or frame. The \emph{Weitzenb\"ock connection} is the Maurer-Cartan form associated to the vielbein
\begin{equation}
W_\mu{}^\vrho{}_\nu = e^\vrho{}_a\,\de_\mu\,e_\nu{}^a,
\end{equation}
which is flat.\footnote{The spin connection corresponding to the Weitzenb\"ock connection is zero (\emph{Weitzenb\"ock gauge}).} Denote by $D_\mu$ the covariant derivative with respect to the Weitzenb\"ock connection. 
The vielbein is covariantly constant:
\begin{equation}
D_\mu\,e_\nu{}^a = \de_\mu\,e_\nu{}^a - W_\mu{}^\vrho{}_\nu\,e_\vrho{}^a = 0.
\end{equation}
The torsion of the Weitzenb\"ock connection or \emph{intrinsic torsion} is the antisymmetric part of the Weitzenb\"ock connection, which is not vanishing in general:\footnote{Here and in the following we denote by $[...]$ and $(...)$ among indices in a tensor components the (anti)symmetrisation \emph{without} numerical coefficient. For example, $A_{[\mu\nu]} = A_{\mu\nu} - A_{\nu\mu}$, $A_{(\mu\nu)} = A_{\mu\nu} + A_{\nu\mu}$.}
\begin{equation}\label{DefTorsion}
T_\mu{}^\vrho{}_\nu = W_{\mu}{}^\vrho{}_{\nu} - W_\nu{}^\vrho{}_\mu = W_{[\mu}{}^\vrho{}_{\nu]}.
\end{equation}
The torsion is a covariant tensor, that is, its variation under diffeomorphisms is generated by the Lie derivative. Indeed, it can be written in a manifestly covariant way:
\begin{equation}\label{TorsionLieDer}
T_\mu{}^\vrho{}_\nu = -e_\mu{}^a\,e_\nu{}^b\,\mathcal{L}_{e_a}e^\vrho{}_b.
\end{equation}
The torsion satisfies the following Bianchi identity
\begin{equation}\label{BianchiId0}
D_{[\mu}\,T_{\nu]}{}^\sigma{}_\vrho 
+ D_\vrho\,T_\mu{}^\sigma{}_\nu 
+ T_\vrho{}^\lambda{}_\mu\,T_\lambda{}^\sigma{}_\nu + T_\vrho{}^\lambda{}_\nu\,T_\mu{}^\sigma{}_\lambda - T_\vrho{}^\sigma{}_\lambda\,T_\mu{}^\lambda{}_\nu  = 0,
\end{equation}
which, introducing the adjoint action
\begin{equation}
\delta_{T_\vrho}\,T_\mu{}^\sigma{}_\nu =
T_\vrho{}^\lambda{}_\mu\,T_\lambda{}^\sigma{}_\nu + T_\vrho{}^\lambda{}_\nu\,T_\mu{}^\sigma{}_\lambda - T_\vrho{}^\sigma{}_\lambda\,T_\mu{}^\lambda{}_\nu,
\end{equation}
can be equivalently written as \cite{Cederwall:2015ica}
\begin{equation}\label{BianchiId}
(D_\vrho + \delta_{T_\vrho})\,T_\mu{}^\sigma{}_\nu = T_\mu{}^\sigma{}_\nu \big{|}_{W_\mu{}^\sigma{}_\nu \rightarrow D_\mu T_\vrho{}^\sigma{}_\nu},
\end{equation}
where in the right-hand side $W_\mu{}^\sigma{}_\nu$ in \eqref{DefTorsion} is replaced by $D_\mu\,T_\vrho{}^\sigma{}_\nu$, with $\vrho$ as a spectator index. The unnecessarily complicated, but equivalent form \eqref{BianchiId} of the Bianchi identity \eqref{BianchiId0} will be convenient in the generalised case. A way to prove \eqref{BianchiId} is to consider the closure condition of the Lie derivative, that is, the Leibniz identity:
\begin{equation}
[\mathcal{L}_\xi,\mathcal{L}_\zeta]\,V^\vrho = \mathcal{L}_{\mathcal{L}_\xi\zeta}\,V^\vrho,
\end{equation}
for any $\xi^\mu,\zeta^\mu,V^\mu$. If one sets $\xi^\mu = e^\mu{}_a\,\tilde{\xi}^a$, $\zeta^\mu = e^\mu{}_a\,\tilde{\zeta}^a$, $V^\mu = e^\mu{}_a\,\tilde{V}^a$, requiring $\tilde{\xi}^a, \tilde{\zeta}^a,\tilde{V}^a$ to be constant, then one can check that
\begin{equation}
([\mathcal{L}_\xi,\mathcal{L}_\zeta]-\mathcal{L}_{\mathcal{L}_\xi\zeta})\,V^\vrho = -\xi^\nu\,\zeta^\tau\,V^\sigma\,[(D_\sigma + \delta_{T_\sigma})\,T_\nu{}^\vrho{}_\tau - D_{[\nu}\,T_{\tau]}{}^\vrho{}_\sigma],
\end{equation}
which is the Bianchi identity \eqref{BianchiId}. 

For a tensor $V^{\mu_1\dots\mu_p}{}_{\nu_1\dots\nu_q}$, defining its ``flattened version" as 
\begin{equation}
\tilde{V}^{a_1\dots a_p}{}_{b_1\dots b_q} = e_{\mu_1}{}^{a_1}\dots e_{\mu_p}{}^{a_p}\,e^{\nu_1}{}_{b_1}\dots e^{\nu_q}{}_{b_q}\,
V^{\mu_1\dots\mu_p}{}_{\nu_1\dots\nu_q},
\end{equation}
one has
\begin{equation}\label{DressingFormula}
D_\mu\,V^{\mu_1\dots\mu_p}{}_{\nu_1\dots\nu_q} =
e^{\mu_1}{}_{a_1}\dots e^{\mu_p}{}_{a_p}\,e_{\nu_1}{}^{b_1}\dots e_{\nu_q}{}^{b_q}\,\de_\mu\, \tilde{V}^{a_1\dots a_p}{}_{b_1\dots b_q}.
\end{equation}

One can show that one of the two traces of the Weitzenb\"ock connection is linked to the derivative of the determinant of the vielbein $e$, according to
\begin{equation}
\de_\mu\,e = e\,W_\mu{}^\nu{}_\nu.
\end{equation}
The formula for the covariant divergence in terms of the Weitzenb\"ock torsion is 
\begin{equation}\label{CovDiv0}
e\,D_\mu\,V^\mu = \de_\mu\,(e\,V^\mu) + e\,T_\mu{}^\mu{}_\nu\,V^\nu,
\end{equation}
where one uses the relation $W_\mu{}^\mu{}_\nu = W_\nu{}^\mu{}_\mu + T_\mu{}^\mu{}_\nu$ between the two traces of the connection.

The Levi-Civita connection differs from the Weitzenb\"ock connection by a combination of the torsion, which is usually called \emph{contorsion} $K_\mu{}^\vrho{}_\nu$:
\begin{equation}\label{DefContorsion}
\Gamma_\mu{}^\vrho{}_\nu = W_\mu{}^\vrho{}_\nu + \tfrac{1}{2}\,(T_{\mu\nu}{}^\vrho + T_{\nu\mu}{}^\vrho - T_\mu{}^\vrho{}_\nu) =: W_\mu{}^\vrho{}_\nu + K_\mu{}^\vrho{}_\nu.
\end{equation}
This relation and the following Bianchi identity for the contorsion
\begin{equation}
D_\mu\,K_\nu{}^\vrho{}_\sigma = \de_\mu\,K_\nu{}^\vrho{}_\sigma - W_\mu{}^\lambda{}_\nu\,K_\lambda{}^\vrho{}_\sigma + W_\mu{}^\vrho{}_\lambda\,K_\nu{}^\lambda{}_\sigma - W_\mu{}^\lambda{}_\sigma\,K_\nu{}^\vrho{}_\lambda,
\end{equation}
which follows from  \eqref{BianchiId0}, allow to express the Riemann tensor in terms of the torsion and of the contorsion:
\begin{equation}
R^\mu{}_{\nu\vrho\sigma} = D_\vrho\,K_\sigma{}^\mu{}_\nu - D_\sigma\,K_\vrho{}^\mu{}_\nu + T_\vrho{}^\tau{}_\sigma\,K_\tau{}^\mu{}_\nu + K_\vrho{}^\mu{}_\tau\,K_\sigma{}^\tau{}_\nu - K_\sigma{}^\mu{}_\tau\,K_\vrho{}^\tau{}_\nu,
\end{equation}
so that in particular the Ricci scalar $\mathscr{R}=R^{\mu\nu}{}_{\mu\nu}$ is
\begin{equation}
\mathscr{R} = -2\,D_\mu\,T^{\mu\nu}{}_\nu  - T_{\mu\nu}{}^\nu\,T^{\mu\vrho}{}_\vrho - \tfrac{1}{4}\,T^{\mu\nu\vrho}\,T_{\mu\nu\vrho} - \tfrac{1}{2}\,T^{\mu\nu\vrho}\,T_{\mu\vrho\nu},
\end{equation}
where the contorsion has been replaced by its definition \eqref{DefContorsion} and the identity $T^{\mu\nu\vrho}\,T_{\nu\mu\vrho} = T^{\mu\nu\vrho}\,T_{\mu\vrho\nu}$ has been used. Moreover, using the formula \eqref{CovDiv0}, one can factor out a total derivative from $e\,\mathscr{R}$: 
\begin{equation}
e\,\mathscr{R} = -\de_\mu\,(2\,e\,T^{\mu\nu}{}_\nu) - e\,(\tfrac{1}{4}\,T^{\mu\nu\vrho}\,T_{\mu\nu\vrho} + \tfrac{1}{2}\,T^{\mu\nu\vrho}\,T_{\mu\vrho\nu} - T_{\mu\nu}{}^\nu\,T^{\mu\vrho}{}_\vrho),
\end{equation}
which shows that, up to a boundary term, the Einstein-Hilbert action admits an equivalent formulation in terms of the quadratic scalar contractions of the torsion:
\begin{equation}\label{HE}
\int\diff^d x\,e\,\mathscr{R} = \int\diff^d x\,e\,\bigg{[}-\frac{1}{4}\,T^{\mu\nu\vrho}\,T_{\mu\nu\vrho} - \frac{1}{2}\,T^{\mu\nu\vrho}\,T_{\mu\vrho\nu} + T_{\mu\nu}{}^\nu\,T^{\mu\vrho}{}_\vrho\bigg{]}.
\end{equation}
This formulation is usually called ``teleparallel equivalent of General Relativity".\footnote{One may notice that the fundamental objects here are the anholonomy coefficients and that interpreting them as the components of the torsion associated with the Weitzenb\"ock connection is not strictly necessary. However the notion of teleparallelism can be appreciated in view of the applications and possible extensions in General Relativity. See for example \cite{Moller1961a, Moller1961b, Moller1978, Hayashi:1979qx}.}

This result can be also obtained by symmetry principles. Since the torsion decomposes into three irreducible components -- the totally antisymmetric part $T_{[\mu\nu\vrho]}$, corresponding to the Young tableau {\ytableausetup{smalltableaux}
$\ydiagram{3}\,$}; the trace $T_{\mu\nu}{}^\nu$, corresponding to {\ytableausetup{smalltableaux}
$\ydiagram{1}\,$}; and the remaining part, corresponding to the traceless mixed Young tableau {\ytableausetup{smalltableaux}
$\ydiagram{2,1}$} -- there are three independent scalar contractions of the torsion, which can be chosen as
\begin{equation}
T^{\mu\nu\vrho}\,T_{\mu\nu\vrho}, \quad T^{\mu\nu\vrho}\,T_{\mu\vrho\nu}, \quad T^{\mu\nu}{}_\nu\,T_{\mu\vrho}{}^\vrho.
\end{equation}
The ansatz for the most general quadratic action is an arbitrary linear combination of the three scalars, multiplied by $e$ and integrated:
\begin{equation}\label{AnsatzAction}
\int\diff^dx\,e\,(\alpha\,T^{\mu\nu\vrho}\,T_{\mu\nu\vrho} + \beta\,T^{\mu\nu\vrho}\,T_{\mu\vrho\nu} + \gamma\,T^{\mu\nu}{}_\nu\,T_{\mu\vrho}{}^\vrho).
\end{equation}
This ansatz is manifestly generally invariant, that is, invariant with respect to diffeomorphisms, since the torsion is a covariant tensor.\footnote{The Lie derivative of a scalar density being a total derivative.} The three coefficients $\alpha,\beta,\gamma$ are fixed by requiring the invariance under local Lorentz transformations, which is not manifest in teleparallel formulation. The Lorentz transformation of the vielbein is
\begin{equation}
\delta_L\,e_\mu{}^a = \tilde{\Lambda}^a{}_b\,e_\mu{}^b,
\end{equation}
where $\tilde{\Lambda}_{ab} = - \tilde{\Lambda}_{ba}$. By direct computation, one gets the transformation of the Weitzenb\"ock connection
\begin{equation}
\delta_L\,W_\mu{}^\vrho{}_\nu = \de_\mu\,\Lambda^\vrho{}_\nu + W_\mu{}^\vrho{}_\sigma\,\Lambda^\sigma{}_\nu - W_\mu{}^\sigma{}_\nu\,\Lambda^\vrho{}_\sigma = D_\mu\,\Lambda^\vrho{}_\nu,
\end{equation}
where $\Lambda^\vrho{}_\nu = e^\vrho{}_a\,e_\nu{}^b\,\tilde{\Lambda}^a{}_b$. Therefore, the transformation of the torsion is 
\begin{equation}
\delta_L\,T_{\mu\nu\vrho} = D_\mu\,\Lambda_{\nu\vrho} - D_\vrho\,\Lambda_{\nu\mu}.
\end{equation}
Using this, one can compute the transformations of the three scalar contractions:
\begin{equations}
\delta_L\,(T^{\mu\nu}{}_\nu\,T_{\mu\vrho}{}^\vrho) &= -2\,T^{\mu\nu}{}_\nu\,D_\vrho\,\Lambda^\vrho{}_\mu,\label{tr1}\\
\delta_L\,(T^{\mu\nu\vrho}\,T_{\mu\nu\vrho}) &= 4\,T^{\mu\nu\vrho}\,D_\mu\,\Lambda_{\nu\vrho},\\
\delta_L\,(T^{\mu\nu\vrho}\,T_{\mu\vrho\nu}) &= -2\,T^{\mu\nu\vrho}\,D_\mu\,\Lambda_{\nu\vrho} -2\,T^{\mu\vrho\nu}\,D_\vrho\,\Lambda_{\nu\mu}.\label{tr3}
\end{equations}
Two terms in the right-hand sides differ by a total derivative:
\begin{equation}\label{IdentityTotDer}
\tfrac{1}{2}\,e\,T^{\mu\vrho\nu}\,D_\vrho\,\Lambda_{\nu\mu} = \de_\mu\,(e\,D_\nu\,\Lambda^{\mu\nu}) + e\,T^{\mu\nu}{}_\nu\,D_\vrho\,\Lambda^\vrho{}_\mu.
\end{equation}
This relation relies on the fact that the torsion is the field strength associated to the Weitzenb\"ock covariant derivative:\footnote{Applying \eqref{CovDiv0} to $V^\mu = D_\nu\,\Lambda^{\mu\nu}$, one gets
\begin{equation}\label{tmp1}
e\,D_\mu\,D_\nu\,\Lambda^{\mu\nu} = \de_\mu\,(e\,D_\nu\,\Lambda^{\mu\nu}) + e\,T_\mu{}^\mu{}_\nu\,D_\vrho\,\Lambda^{\nu\vrho},
\end{equation}
where in the left-hand side the antisymmetric part $D_\mu\,D_\nu\,\Lambda^{\mu\nu} = \tfrac{1}{2}\,[D_\mu,D_\nu]\,\Lambda^{\mu\nu}$ is selected. On the other hand, using the formula \eqref{DressingFormula} twice, one gets
\begin{align}
D_\mu\,D_\nu\,\Lambda^{\vrho\sigma} = -W_\mu{}^\lambda{}_\nu\,D_\lambda\,\Lambda^{\vrho\sigma} + e^\vrho{}_a\,e^\sigma{}_b\,\de_\mu\,\de_\nu\,\Lambda^{ab}.\label{DoubleCovDer}
\end{align}
Taking the antisymmetric part one gets the expression \eqref{CommCovDer}. Finally tracing this formula and replacing in \eqref{tmp1}, one gets \eqref{IdentityTotDer}.}
\begin{equation}\label{CommCovDer}
[D_\mu,D_\nu]\,\Lambda^{\vrho\sigma} =  -T_\mu{}^\lambda{}_\nu\,D_\lambda\,\Lambda^{\vrho\sigma}.
\end{equation}
Using the transformations \eqref{tr1}--\eqref{tr3} and the relation \eqref{IdentityTotDer}, one can compute the transformation of the ansatz \eqref{AnsatzAction}: 
\begin{align}
& \delta_L\,(\alpha\,e\,T^{\mu\nu\vrho}\,T_{\mu\nu\vrho} + \beta\,e\,T^{\mu\nu\vrho}\,T_{\mu\vrho\nu} + \gamma\,e\,T^{\mu\nu}{}_\nu\,T_{\mu\vrho}{}^\vrho) \nn\\
=&\; 2\,(2\alpha-\beta)\,e\,T^{\mu\nu\vrho}\,D_\mu\,\Omega_{\nu\vrho} - (2\,\beta + \gamma)\,e\,T^{\mu\nu\vrho}\,D_\vrho\,\Omega_{\nu\mu} + \gamma\,\de_\mu(e\,D_\nu\,\Omega^{\mu\nu}).
\end{align}
where one uses $\delta_L\,e = 0$ since $\tilde{\Lambda}$ has vanishing determinant.
Therefore, up to boundary terms, the ansatz \eqref{AnsatzAction} is Lorentz-invariant if and only if
\begin{equation}
\beta = 2\,\alpha, \quad \gamma = -4\,\alpha.
\end{equation}
Choosing $\alpha = \tfrac{1}{4}$ we get precisely the combination in \eqref{HE} equivalent to the Einstein-Hilbert action.
As it will be shown, in generalised geometry the torsion is not simply the antisymmetric part of the Weitzenb\"ock connection, so that the relation \eqref{CommCovDer} is not valid anymore. Nevertheless, as shown in \cite{Cederwall:2021xqi}, one could equivalently integrate by parts in the Lorentz variation of the ansatz of the action, and then use the Bianchi identity. This is the way which can be naturally extended to the case of generalised geometry. In the next section we will apply this strategy in the case of the $n=8$ exceptional geometry.

\section{$E_{8(8)}$ exceptional geometry}\label{Sec2}

In this section we fix the notation for the $n=8$ exceptional geometry and we collect some useful identities. The full duality group is $E_{8(8)} \times \mathbb{R}^+$, where $E_{8(8)}$ denotes the maximally non-compact form of the exceptional group, and $\mathbb{R}^+$ is the trombone duality", that is, the invariance of the equations of motion under a rescaling \cite{Cremmer:1997xj, LeDiffon:2008sh}.

The generalised Lie derivative generated by a vector $\xi^M$ and an ancillary parameter $\sigma_M$ of a vector density $V_{[\lambda]}^M$ with weight $\lambda$ is defined by \cite{Hohm:2014fxa}\footnote{Here the weight of a vector density with respect to the generalised Lie derivative is defined as the opposite of the coefficient of the term $\de_N X^N$ where the $Y$ tensor is replaced.}
\begin{align}
L_{\xi,\sigma}\,V_{[\lambda]}^M =&\, \xi^N\,\de_N\,V_{[\lambda]}^M - V_{[\lambda]}^N\,\de_N\,\xi^M -(\lambda + 1)\,\de_N \xi^N\,V_{[\lambda]}^M + Y^{MP}{}_{QN}\,\de_P\,\xi^Q\,\,V_{[\lambda]}^N + f^{MR}{}_N\,\sigma_R\,V_{[\lambda]}^N \nn\\
=&\, \xi^N\,\de_N\,V_{[\lambda]}^M - \lambda\,\de_N \xi^N\,V_{[\lambda]}^M - f^{RM}{}_N\,f_R{}^P{}_Q\,\de_P\,\xi^Q\,V_{[\lambda]}^N+ f^{MR}{}_N\,\sigma_R\,V_{[\lambda]}^N,
\end{align}
where
\begin{equation}
Y^{MP}{}_{QN} = \delta^M_Q\,\delta^P_N + \delta^M_N\,\delta^P_Q - f^{RM}{}_N\,f_R{}^P{}_Q,
\end{equation}
where $f_{MNP}$ are the structure constants of the algebra of $E_{8(8)}$. Similarly, for a density one-form $U^{[\lambda]}_M$ of density $\lambda$,
\begin{align}
L_{\xi,\sigma}\,U^{[\lambda]}_M =&\, \xi^N\,\de_N\,U^{[\lambda]}_M + U^{[\lambda]}_N\,\de_M\,\xi^M +(\lambda + 1)\,\de_N\,\xi^N\,U^{[\lambda]}_M - Y^{NP}{}_{QM}\,\de_P\,\xi^Q\,U^{[\lambda]}_N - f^{NR}{}_M\,\sigma_R\,U^{[\lambda]}_N \nn\\
=&\, \xi^N\,\de_N\,U^{[\lambda]}_M + \lambda\,\de_N\,\xi^N\,U^{[\lambda]}_M + f^{RN}{}_M\,f_R{}^P{}_Q\,\de_P\,\xi^Q\,U^{[\lambda]}_N - f^{NR}{}_M\,\sigma_R\,U^{[\lambda]}_N.
\end{align}
We usually consider vector densities of weight $\lambda = -1$ and density one-forms of weight $\lambda = 1$, and in these cases we omit the decoration, writing simply  $V^M_{[-1]} =: V^M$ and $U_M^{[1]} =: U_M$.

The ancillary parameters are necessary for the generalised Lie derivative to close into an algebra, that is, the commutator of two generalised Lie derivatives is still a generalised Lie derivative, provided that the \emph{section constraint} holds \cite{Hohm:2014fxa}
\begin{equation}\label{SectionConstraintY}
Y^{MP}{}_{QN}\,A_M\,B_P =
A_Q\,B_N + A_N\,B_Q - f^{RM}{}_N\,f_R{}^P{}_Q\,A_M\,B_P \approx 0,
\end{equation}
where $A_M, B_M$ are partial derivatives with respect to the generalised coordinates $\de_M$ (with any argument), or ancillary parameters $\sigma_M$ in the generalised Lie derivative.\footnote{This means
\begin{equations}
& Y^{MP}{}_{QN}\,\de_M(\cdot)\,\de_P(\cdot) \approx 0, \quad
Y^{MP}{}_{QN}\,\de_M\,\de_P(\cdot) \approx 0, \\
& Y^{MP}{}_{QN}\,\sigma_M\,\de_P(\cdot) \approx 0, \quad
Y^{MP}{}_{QN}\,\de_M(\cdot)\,\sigma_P \approx 0, \\
& Y^{MP}{}_{QN}\,\sigma_M\,\tau_P \approx 0,
\end{equations}
for any ancillary parameter $\sigma_M$, $\tau_M$.}
We will denote by $\approx$ the equality modulo the section constraint. The section constraint corresponds to set to zero the projections on the irreducible components 
$\bm{1}$, $\bm{248}$, and $\bm{3875}$ of the algebra of $E_{8(8)}$, as one can see by using the Jacobi identity \eqref{DecJacobi}:
\begin{equation}\label{SecCon}
{P^{(n)}}_{MN}{}^{PQ}\,A_P\,B_Q \approx 0, \quad n=1, 248, 3875,
\end{equation}
where $P^{(n)}$ are the projectors in \eqref{P1}--\eqref{P3875}, that is, 
\begin{equations}
& \eta^{MN}\,A_M{}\,B_N \approx 0,\\
& f^{MNP}\,A_M\,B_N \approx 0,\\
& f_{MN}{}^{PQ}\,A_P\,B_Q \approx 2\,(A_M\,B_N + A_N\,B_M),
\end{equations}
where $\eta_{MN}$ is the Cartan-Killing metric 
\begin{equation}
f^{MN}{}_P\,f_{MNQ} = -60\,\eta_{PQ},
\end{equation}
and $f^{MP}{}_{NQ}$ is the symmetrised product of two structure constants
\begin{equation}
f^{MP}{}_{NQ} := f^{RM}{}_N\,f_R{}^P{}_Q + f^{RP}{}_N\,f_R{}^M{}_Q.
\end{equation}

The closure condition of the generalised Lie derivative, assuming the section constraint, is \cite{Hohm:2014fxa} (see also \cite{Rovere:2025jmp})
\begin{equation}
[L_{\xi,\sigma},L_{\eta,\tau}] \approx L_{\frac{1}{2}\,(L_{\xi}\eta-L_{\eta}\xi),L_{\xi}\tau-L_{\eta}\sigma+\sigma(\xi,\eta)},
\end{equation}
where $L_\xi = L_{\xi,0}$ is the generalised Lie derivative with vanishing ancillary parameter, and $\sigma(\xi,\eta)$ denotes the combination
\begin{equation}
\sigma_M(\xi,\eta)=\tfrac{1}{2}\,f^P{}_{RS}\,(\xi^R\,\de_P\,\de_M\,\eta^S-\eta^R\,\de_P\,\de_M\,\xi^S),
\end{equation}
which can be used as an ancillary parameter since it is on section. Observe also that, if $\sigma_M$ is on section, $L_\xi\,\sigma_M$ is on section too.

Consider a \emph{frame} $E^M{}_A$ in  the duality group $E_{8(8)} \times \mathbb{R}^+$. Denote its inverse (\emph{vielbein}) with $E_M{}^A$, its $E_{8(8)}$ component with $U^M{}_A$, and its $\mathbb{R}^+$ component with $r^{-1}$, so that \cite{Inverso:2024xok} (see also \cite{Rovere:2025jmp})
\begin{equation}\label{EUR}
E^M{}_A = r^{-1}\,U^M{}_A, \quad
E_M{}^A = r\,U_M{}^A,
\end{equation}
where $U_M{}^A$ is the inverse of $U^M{}_A$. The $E_{8(8)}$ invariance of $\eta_{MN}$ and $f_{MNP}$ reads
\begin{equation}
U^M{}_A\,U^N{}_B\,\eta_{MN} = \eta_{AB}, \quad
U^M{}_A\,U^N{}_B\,U^P{}_C\,f_{MNP} = f_{ABC}.
\end{equation}

Define the following matrix
\begin{equation}\label{Mmatrix}
M_{MN} = U_M{}^A\,U_N{}^B\,\delta_{AB},
\end{equation}
using the previous properties \eqref{propDelta1}--\eqref{propDelta3}, one finds the following relations:
\begin{equations}
& M_N{}^N = M_{MN}\,\eta^{MN} = 8, \label{RelM}\\
& M_{MN}\,M^{MN} = 248, \\
& M^{MN}\,f_{MN}{}^{PQ} = 4\,(M^{PQ}-\eta^{PQ}), \label{FsymM}\\
& M^{MP}\,M^{NQ}\,M^{LR}\,f_{MNL} = -f^{PQR}.\label{RelMlast}
\end{equations}

The \emph{Weitzenb\"ock connection} associated with the frame $E$ is defined as in ordinary geometry \cite{Coimbra:2011ky, Coimbra:2012af, Inverso:2017lrz}:\footnote{Using $\de_M\,E^N{}_A = - E^P{}_A\,E^N{}_B\,\de_M\,E_P{}^B$, it reads also $W_M{}^P{}_N = - E_N{}^A\,\de_M\,E^P{}_A$.}
\begin{equation}\label{DefW}
W_M{}^P{}_N = E^P{}_A\,\de_M\,E_N{}^A =
U^P{}_A\,\de_M\,U_N{}^A + \de_M\,\text{log}\,r\,\delta^P_N.
\end{equation}
It is useful to introduce the following parametrisation:
\begin{equation}\label{DecompositionW}
W_M{}^P{}_N = \tfrac{1}{2}\,W'_M\,\delta^P_N - W''_{ML}\,f^{LP}{}_N, 
\end{equation}
where $W'_M$ captures the $\mathbb{R}^+$ part and $W''_{MN}$ the $E_{8(8)}$ part:
\begin{equations}
& W'_M = \tfrac{1}{124}\,W_M{}^N{}_N = 2\,\de_M\log{r}, \\
& W''_{MN} = \tfrac{1}{60}\,f_{NP}{}^Q\,W_M{}^P{}_Q = \tfrac{1}{60}\,f_{NP}{}^Q\,U^P{}_A\,\de_M\,U_Q{}^A.
\end{equations}

The \emph{generalised intrinsic torsion} is defined starting from its definition in terms of the Lie derivative in ordinary geometry \eqref{TorsionLieDer}, with the Lie derivative replaced by the generalised one, which ensures its covariance \cite{Coimbra:2011ky, Coimbra:2012af, Inverso:2017lrz, Inverso:2024xok}:
\begin{align}
T_M{}^P{}_N &= - E_M{}^A\, E_N{}^B\, L_{E_A,h_A} E^P{}_B \nn\\
&= W_{[M}{}^P{}_{N]} + Y^{PR}{}_{QN}\,W_R{}^Q{}_M + f^{PR}{}_N\,E_M{}^A\,h_{RA} \nn\\
&= W_M{}^P{}_N + W_S{}^S{}_M\,\delta^P_N + f^{PQ}{}_N\,f_Q{}^R{}_S\,W_R{}^S{}_M + f^{PR}{}_N\,E_M{}^A\,h_{RA},\label{GenTorsionW}
\end{align}
where the second line shows the deformation with respect to ordinary geometry, parametrised by the $Y$ tensor and by the term depending on $h_{MA}$. The latter is on section in its first index and it is fixed in such a way that the torsion contains only the irreducible representations $\bf{1}\oplus\bf{248}\oplus\bf{3875}$ (see Appendix \ref{AppDiff}), which are the ones of the embedding tensor components in 3d gauged supergravity. The right choice is
\begin{equation}
E_M{}^A\,h_{RA} = \tfrac{1}{60}\,f_{MS}{}^Q\,W_R{}^S{}_Q,
\end{equation}
so that the full expression of the torsion is \cite{Inverso:2024xok}
\begin{equation}
T_M{}^P{}_N = W_M{}^P{}_N + W_S{}^S{}_M\,\delta^P_N + f^{PQ}{}_N\,f_Q{}^R{}_S\,W_R{}^S{}_M + \tfrac{1}{60}\,f^{PR}{}_N\,f_{MS}{}^Q\,W_R{}^S{}_Q.
\end{equation}
The fact that the generalised intrinsic torsion sits in the representation $\bf{1}\oplus\bf{248}\oplus\bf{3875}$ allows for the following parametrisation \cite{Inverso:2017lrz}:
\begin{equation}\label{DecTor}
T_M{}^P{}_N = \vtheta'_M\,\delta^P_N + \tfrac{1}{2}\,f^{QP}{}_N\,f_{QM}{}^R\,\vtheta'_R - f^{QP}{}_N\,\vtheta''_{MQ}  - f_M{}^Q{}_N\,\vtheta,
\end{equation}
where $\vtheta$, $\vtheta'_M$, and $\vtheta''_{MN}$ are the components in the representations $\bf{1}$, $\bf{248}$, and $\bf{3875}$ respectively. In particular, $\vtheta''_{MN}$ is symmetric traceless, such that $P^{(3875)}{}_{MN}{}^{PQ}\,\vtheta''_{PQ} = \vtheta''_{MN}$, or
\begin{equation}
f_{MN}{}^{PQ}\,\vtheta_{PQ} = -24\,\vtheta''_{MN}.
\end{equation} 
The same information of the torsion is carried by its dual with respect to the structure constants, which are denoted with $T''_{ML}$:
\begin{equation}
T''_{ML} = \tfrac{1}{60}\,T_M{}^P{}_N\,f_{LP}{}^N = \tfrac{1}{2}\,f_{ML}^R\,\vtheta'_R + \vtheta''_{ML} + \eta_{ML}\,\vtheta.
\end{equation}
Its irreducible components correspond to its trace, its antisymmetric part, and its traceless symmetric part:\footnote{The traces of the torsion are $T_M{}^M{}_N = - 29\,\vtheta'_N$, $T_M{}^{PM} = 31\,\vtheta'^P$, $T_M{}^N{}_N = 248\,\vtheta'_M$.}
\begin{equations}
& \vtheta = \tfrac{1}{248}\,T''_M{}^M = \tfrac{1}{248}\tfrac{1}{60}\,T_M{}^P{}_N\,f^M{}_P{}^N, \\
& \vtheta'_N = -\tfrac{1}{30}\,f_N{}^{ML}{}\,T''_{ML}, \\
& \vtheta''_{ML} + \eta_{ML}\,\vtheta = \tfrac{1}{2}\,(T''_{ML} + T''_{LM}).
\end{equations}
The components of the torsion satisfy the following identities involving the matrix \eqref{Mmatrix}:
\begin{equation}\label{GroupThM1} 
M^{MN}\,\vtheta''_{MN} = 0,\quad
M^{PQ}\,f_P{}^{RM}\,(\vtheta'_R-W'_R)\,\vtheta''_{MQ} = 0.
\end{equation}
and they can be written in terms of the components of the Weitzenb\"ock connection, defined in \eqref{DecompositionW}:
\begin{equations}
& \vtheta = \tfrac{1}{4}\,W''_M{}^M, \label{theta0}\\
& \vtheta'_M = W'_M + f_M{}^{PQ}{}\,W''_{PQ},\label{theta1}\\
& \vtheta''_{MN} + \eta_{MN}\,\vtheta =
W''_{MN} + W''_{NM} - \tfrac{1}{2}\,f_{MN}{}^{PQ}\,W''_{PQ}.\label{theta2}
\end{equations}
Equivalently, $\vtheta$, $\vtheta'_M - W'_M$, and $\vtheta''_{MN}$ are the projections of $W''_{MN}$ in the representations $\bf{1}, \bf{248}, \bf{3875}$ respectively:
\begin{equations}
& {P^{(1)}}_{MN}{}^{PQ}\,W''_{PQ} = \tfrac{1}{62}\,\eta_{MN}\,\vtheta, \label{thetaPW1}\\
& {P^{(248)}}_{MN}{}^{PQ}\,W''_{PQ} =  -\tfrac{1}{60}\,f_{MN}{}^P\,(\vtheta'_P - W'_P), \label{thetaPW2}\\
& {P^{(3875)}}_{MN}{}^{PQ}\,W''_{PQ} = \tfrac{1}{14}\,\vtheta''_{MN},\label{thetaPW3}
\end{equations}
so that the full decomposition of $W''_{MN}$ in irreducible components reads
\begin{equation}
W''_{MN} = \tfrac{1}{62}\,\eta_{MN}\,\vtheta - \tfrac{1}{60}\,f_{MN}{}^P\,(\vtheta'_P-W'_P) + \tfrac{1}{14}\,\vtheta''_{MN} + W_{MN}^{(27000)} + W_{MN}^{(30380)}.
\end{equation}
The last two components, which satisfy
\begin{equation}
f_{MN}{}^{PQ}\,W_{MN}^{(27000)} = 4\,W_{MN}^{(27000)},\quad
f^{MNP}\,W_{MN}^{(30380)} =
f^{MPNQ}\,W_{MN}^{(30380)} = 0,
\end{equation}
sit in the representation $\bm{27000}$ and $\bm{30380}$ respectively, since
\begin{equation}
\bm{248} \otimes \bm {248} = \bm{1} \oplus \bm{248} \oplus \bm{3875} \oplus \bm{27000} \oplus \bm{30380}.
\end{equation}

The covariant derivative $D_M$ with respect to the Weitzenb\"ock connection of $E_M{}^A$ is defined in such a way that $E_M{}^A$ is covariantly constant
\begin{align}
0 &\equiv D_M^{(-1/2)}\,E_N{}^A = \de_M\,E_N{}^A - W_M{}^P{}_N\,E_P{}^A \nn\\
&= \de_M\,E_N{}^A + W''_{MP}\,f^{PL}{}_N\,E_L{}^A - \tfrac{1}{2}\,W'_M\,E_N{}^A,
\label{DerWE}
\end{align}
where $n$ in $D^{(n)}_M$ denotes the weight with respect to $W'_M$. The previous condition is an identity thanks to the definition of the Weitzenb\"ock connection in \eqref{DefW}. As a consequence, $r$, which is equal to the power $|E|^{1/248}$ of the determinant of $E_M{}^A$, is covariantly constant too
\begin{equation}
D_M^{({-1/2})}\,r = 0.
\end{equation} 
The covariant derivative of a vector density $V_{[\lambda]}^M$ is consistently defined as
\begin{align}
D_M^{(1/2+\lambda)}\,V_{[\lambda]}^N &:= \de_M\,V_{[\lambda]}^N + W_M{}^N{}_L\,V_{[\lambda]}^L + \tfrac{\lambda}{124}\,W_M{}^L{}_L\,V_{[\lambda]}^N \nn\\
&= \de_M\,V_{[\lambda]}^N - W''_{MP}\,f^{PN}{}_L\,V_{[\lambda]}^L + (\tfrac{1}{2}+\lambda)\,W'_M\,V_{[\lambda]}^N.
\end{align}
In the case of generalised vectors $V^M$ of weight $\lambda = -1$, one has the following identity, which will be useful in the following\footnote{More in general, 
\begin{align}
\de_M\,(r^\beta\,V^M_{[\lambda]}) &= r^\beta\,(\de_M\,V_{[\lambda]}^M + \beta\,\de_M\,\log{r}\,V_{[\lambda]}^M) \nn\\
&= r^\beta\,(D^{(1/2+\lambda)}_M\,V_{[\lambda]}^M + W''_{MP}\,f^{PM}{}_L\,V_{[\lambda]}^L - (\tfrac{1}{2}+\lambda)\,W'_M\,V_{[\lambda]}^M + \tfrac{\beta}{2}\,W'_M\,V_{[\lambda]}^M) \nn\\
&= r^\beta\,[D^{(1/2+\lambda)}_M - (W''_{NP}\,f^{NP}{}_M + \tfrac{1+2\,\lambda -\beta}{2}\,W'_M)]\,V_{[\lambda]}^M \nn\\
& \equiv r^\beta\,(D^{(1/2+\lambda)}_M-\vtheta'_M)\,V_{[\lambda]}^M
\Leftrightarrow \beta = 2\,\lambda -1.
\end{align}}
\begin{equation}\label{CovDiv}
\de_M\,(r^{-3}\,V^M) = r^{-3}\,(D^{(-1/2)}_M - \vtheta'_M)\,V^M.
\end{equation}

Similarly to the ordinary geometry case, the generalised torsion satisfies on section a \emph{Bianchi identity}, which follows from the closure condition (Leibniz identity) of the generalised Lie derivative. Such identity reads 
\begin{equation}\label{BianchiT}
(D^{(-1/2)}_M + \delta_{T_M})\,T_N{}^P{}_Q \approx T_N{}^P{}_Q \big{|}_{W_N{}^P{}_Q \rightarrow D^{(-1/2)}_N\,T_M{}^P{}_Q},
\end{equation}
(compare with \eqref{BianchiId}), where 
\begin{equation}
\delta_{T_M}\,T_N{}^P{}_Q = T_M{}^R{}_N\,T_R{}^P{}_Q + T_M{}^R{}_Q\,T_N{}^P{}_R - T_M{}^P{}_R\,T_N{}^R{}_Q.
\end{equation}
Explicitly,
\begin{align}\
& D^{(-1/2)}_M\,T_N{}^P{}_Q + T_M{}^R{}_Q\,T_N{}^P{}_R - T_M{}^R{}_Q\,T_M{}^P{}_R + T_M{}^R{}_M\,T_R{}^P{}_Q \nn\\
\approx &\, D^{(-1/2)}_N\,T_M{}^P{}_Q - D^{(-1/2)}_Q\,T_M{}^P{}_N + Y^{PM}{}_{SQ}\,D^{(-1/2)}_R\,T_M{}^S{}_N \nn\\
&\,+ \tfrac{1}{60}\,f^{RP}{}_Q\,f_{NS}{}^T\,D^{(-1/2)}_R\,T_M{}^S{}_T.\label{BianchiExpl}
\end{align}
When the torsion is covariantly constant, the Bianchi identity reduces to the analogue of the quadratic constraint of the embedding tensor 
\begin{equation}
0 = \delta_{T_M}\,T_N{}^P{}_Q \approx -[T_M,T_N]^P{}_Q + T_M{}^R{}_N\,T_R{}^P{}_Q,
\end{equation}
where $(T_M)^P{}_Q = T_M{}^P{}_Q$.

Starting from the Bianchi identity \eqref{BianchiExpl}, one can derive the following ones (see Appendix \ref{App3} for details):
\begin{equations}
D^{(-1/2)}_M\,\vtheta'^M &\approx \vtheta'_M\,\vtheta'^M, \label{Bianchi4}\\
f^{MN}{}_L\,D^{(-1/2)}_M\,\vtheta'_N &\approx -8\,D^{(-1/2)}_L\,\vtheta + \vtheta'^M\,\vtheta'''_{ML} + \vtheta\,\vtheta'_L, \label{Bianchi1}\\
D^{(-1/2)}_M\,\vtheta''^M{}_L &\approx 7\,D^{(-1/2)}_L\,\vtheta - \tfrac{1}{2}\,\vtheta'^M\,\vtheta''_{ML} + \tfrac{7}{2}\,\vtheta\,\vtheta'_L,\label{Bianchi2}\\
f^{LR}{}_{[M}\,(D^{(-1/2)}_R-\vtheta'_R)\,\vtheta''_{N]L} &\approx 3\,D^{(-1/2)}_{[M}\,\vtheta'_{N]} + 2\,f_{MN}{}^L\,D^{(-1/2)}_L\,\vtheta \nn\\
& + \tfrac{3}{2}\,f_{MN}{}^L\,\vtheta\,\vtheta'_L - \tfrac{1}{2}\,f_{MN}{}^R\,\vtheta'^L\,\vtheta''_{LR}. \label{Bianchi3}
\end{equations}
When the torsion is covariantly constant, one consistently gets the analogue of the quadratic constraints for the embedding tensor \cite{LeDiffon:2008sh}:
\begin{equation}\label{QuadraticConstr}
\vtheta'^M\,\vtheta''_{ML} \approx 0, \quad
\vtheta\,\vtheta'_L \approx 0, \quad
\vtheta'_M\,\vtheta'^M \approx 0.
\end{equation}

\section{$E_8$ ExFT potential in teleparallel formulation}\label{Sec3}

In order to derive the $n=8$ ExFT potential  $V$, first found in \cite{Hohm:2014fxa} in terms of derivatives of $M_{MN}$, using the teleparallel approach introduced in generalised geometry in \cite{Cederwall:2021xqi} (see also \cite{Cederwall:2023xbj,Bossard:2024gry}), one considers an arbitrary combination of the six possible independent quadratic scalar contractions of the components of the torsion, involving $\eta_{MN}$ and $M_{MN}$:
\begin{align}
V = r^{-3} &\,[\alpha\,\vtheta^2 + \beta\,\vtheta'_M\,\vtheta'^M + 
\gamma\,\vtheta''_{MN}\,\vtheta''^{MN} + \vepsilon\,\vtheta'_M\,\vtheta'_N\,M^{MN} \nn\\
+&\, \zeta\,\vtheta''_{MP}\,\vtheta''_{NQ}\,M^{MN}\,M^{PQ} + \xi\,\vtheta''_{MP}\,\vtheta''^M{}_Q\,M^{PQ}].\label{AnsatzV}
\end{align}
As in the case of ordinary geometry, where the Einstein-Hilbert action is recovered by considering the most general combination of the quadratic scalar contractions of the torsion and imposing the Lorentz-invariance, here it is only needed to impose the invariance under the $\text{SO}(16)$ infinitesimal transformation -- denote it with $\delta_L$ -- which acts on $U_M{}^A$ as a rotation and preserves $r$: 
\begin{equation}
\delta_L\,U_M{}^A = \Omega^A{}_B\,U_M{}^B,
\quad \delta_L\,r = 0,
\end{equation}
where $\Omega_{AB}$ is antisymmetric. The Cartan-Killing metric and the structure constants are left invariant, so that the matrix $M$ is invariant too:
\begin{equation}
\delta_L\,\eta_{MN} = \delta_L\,f_{MNP} = \delta_L \,\delta_{AB} = \delta_L\,M_{MN} = 0.
\end{equation}
Using its definition \eqref{DefW} and \eqref{EUR}, one can compute the variation of the Weitzenb\"ock connection
\begin{equation}
\delta_L\,W_M{}^P{}_N = D^{(-1/2)}_M\,\Lambda^P{}_N,
\end{equation}
where $\Lambda^P{}_Q := U^P{}_A\,U_Q{}^B\,\Omega^A{}_B$. Using the decomposition of the Weitzenb\"ock connection \eqref{DecompositionW},
\begin{equation}
\delta_L\,W'_M = 0, \quad
\delta_L\,W''_{MN} = D^{(-1/2)}_M\,\tilde{\Lambda}_N, 
\end{equation}
where
\begin{equation}\label{LambdaTilde}
\tilde{\Lambda}_M := \tfrac{1}{60}\,f_{MP}{}^Q\,\Lambda^P{}_Q.
\end{equation}

In the computation of the variation of the ansatz of the potential \eqref{AnsatzV}, one needs to integrate by parts in order to use the Bianchi identities \eqref{Bianchi1}--\eqref{Bianchi3}. The relevant relation, which follows from \eqref{CovDiv}, is the following one:
\begin{equation}\label{IntByParts}
r^{-3}\,V^M{}_N\,D^{(-1/2)}_M\,\tilde{\Lambda}^N = \de_M\,(r^{-3}\,V^M{}_N\,\tilde{\Lambda}^N) - r^{-3}(D^{(-1/2)}_M-\vtheta'_M)\,V^M{}_N\,\tilde{\Lambda}^N,
\end{equation}
for any $V^M{}_N$. 

The variation of the first five invariants in \eqref{AnsatzV} can be directly computed
by using the following relations, which follow from \eqref{thetaPW1}--\eqref{thetaPW3}:
\begin{equations}
& \vtheta^2 = \tfrac{31}{2}\,{P^{(1)}}_{MN}{}^{PQ}\,W''_{PQ}\,W''^{MN}, \label{Inv1}\\
& (\vtheta'_M-W'_M)\,(\vtheta'^M-W'^M) = -60\,{P^{(248)}}_{MN}{}^{PQ}\,W''_{PQ}\,W''^{MN}, \label{Inv2}\\
& \vtheta''_{MN}\,\vtheta''^{MN} = 196\,{P^{(3875)}}_{MN}{}^{PQ}\,W''_{PQ}\,W''^{MN},\label{Inv3}\\
& (\vtheta'_M-W'_M)\,(\vtheta'_N-W'_N)\,M^{MN} = 60\,{P^{(248)}}_{MNKL}\,M^{PK}\,W^{LQ}\,W''_{PQ}\,W''^{MN} \label{Inv3},\\
& \vtheta''_{MP}\,\vtheta''_{NQ}\,M^{MN}\,M^{PQ} = 196\,{P^{(3875)}}_{MNKL}\,M^{PK}\,M^{LQ}\,W''_{PQ}\,W''^{MN}.
\end{equations}

The variation of the last invariant can be more easily computed by observing that it can be written in this way
\begin{align}
\vtheta''_{MP}\,\vtheta''^M{}_Q\,M^{PQ} &= 
\eta^{MN}\,M^{PQ}\,(W''_{MP}+W''_{PM} - f_M{}^{SR}\,g_P{}^T{}_R\,W''_{ST})\\
&\times (W''_{NQ}+W''_{QN}-f_N{}^{VU}\,f_Q{}^X{}_U\,W''_{UX}) -8\,\vtheta^2 \\
& - \tfrac{1}{2}\,M^{PQ}\,(\vtheta'_P-W'_P)\,(\vtheta'_Q-W'_Q) + \tfrac{1}{2}\,(\vtheta'_M-W'_M)\,(\vtheta'^M-W'^M),\label{RelLastInvariant}
\end{align}
where the first line can be written in terms of the components of the torsion, by using \eqref{GroupThM1} and the following expression:\footnote{Using \eqref{DecJacobi} and \eqref{theta0}--\eqref{theta2},
\begin{align}
W''_{MP}+W''_{PM} - f_M{}^{SR}\,f_P{}^T{}_R\,W''_{ST} &= W''_{MP} + W''_{PM} - \tfrac{1}{2}\,f_{MP}{}^{ST}\,W''_{ST}-\tfrac{1}{2}\,f_{MPR}\,f^{STR}\,W''_{ST} \\
&= \vtheta''_{MP} + \eta_{MP}\,\vtheta - \tfrac{1}{2}\,f_{MP}{}^R\,(\vtheta'_R-W'_R).
\end{align}
}
\begin{equation}
W''_{MP} + W''_{PM} - f_M{}^{SR}\,g_P{}^T{}_R\,W''_{ST} = \vtheta''_{MP} + \eta_{MP}\,\vtheta - \tfrac{1}{2}\,f_{MPR}\,(\vtheta'_R - W'_R).
\end{equation}
The difficult part in computing the variation of \eqref{RelLastInvariant} is
\begin{align}
& \delta_L\,[\tfrac{r^{-3}}{2}\,\eta^{MN}\,M^{PQ}\,(W''_{MP}+W''_{PM} - f_M{}^{SR}\,g_P{}^T{}_R\,W''_{ST})\\
&\times (W''_{NQ}+W''_{QN}-f_N{}^{VU}\,f_Q{}^X{}_U\,W''_{UX})] = \nn\\
=\,& r^{-3}\,\eta^{MN}\,M^{PQ}\,(W''_{(MP)} - f_M{}^{SR}\,g_P{}^T{}_R\,W''_{ST})\\
&\times (D^{(-1/2)}_N\,\tilde{\Lambda}_Q+D^{(-1/2)}_Q\,\tilde{\Lambda}_L-f_N{}^{VU}\,f_Q{}^X{}_U\,D^{(-1/2)}_U\,\tilde{\Lambda}_X)\\
=\,& r^{-3}\,\eta^{MN}\,M^{PQ}\,(\vtheta''_{MP} + \eta_{MP}\,\vtheta - \tfrac{1}{2}\,f_{MPR}\,(\vtheta'_R - W'_R))\\
&\times (D^{(-1/2)}_N\,\tilde{\Lambda}_Q+D^{(-1/2)}_Q\,\tilde{\Lambda}_L-f_N{}^{VU}\,f_Q{}^X{}_U\,D^{(-1/2)}_U\,\tilde{\Lambda}_X).
\end{align}
Integrating by parts, using \eqref{IntByParts}, one gets, up to a total derivative,
\begin{align}
& -r^{-3}\,\eta^{MN}\,M^{PQ}\,[(D^{(-1/2)}_U-\vtheta'_U)\,\vtheta''_{MP} + \eta_{MP}\,(D^{(-1/2)}_U-\vtheta'_U)\,\vtheta \\
&\qquad\qquad\qquad\quad - \tfrac{1}{2}\,f_{MPR}\,(D^{(-1/2)}_U-\vtheta'_U)\,(\vtheta'_R - W'_R)](\delta^U_N\,\tilde{\Lambda}_Q+\delta^U_Q\,\tilde{\Lambda}_L-f_N{}^{VU}\,f_Q{}^X{}_U\,\tilde{\Lambda}_X).
\end{align}

This expression can be simplified by using the following properties, which follow from \eqref{RelM}, \eqref{FsymM}, \eqref{LambdaTilde}, and from the fact that $\Lambda_{MN}$ is antisymmetric:
\begin{equation}
M^{MN}\,\tilde{\Lambda}_N = -\tilde{\Lambda}^M, \quad
f_{PMN}\,M^{(\Lambda)PM} = 4\,\tilde{\Lambda}_N,
\end{equation}
where one defines $M^{(\Lambda)}_{PU} := M_{PQ}\,f^{QX}{}_U\,\tilde{\Lambda}_X$, which turns out to be antisymmetric. Moreover, one has to use the Bianchi identities \eqref{Bianchi1}--\eqref{Bianchi3}, which imply also
\begin{align}
(D^{(-1/2)}_V-\vtheta'_V)\,\vtheta''_{MP}\,f^{MV}{}_U\,M^{(\Lambda)PU} =&\, 
-3\,D^{(-1/2)}_P\,\vtheta'_U\,M^{(\Lambda)PU} - 4\,\tilde{\Lambda}^L\,D^{(-1/2)}_L\,\vtheta \\
&- 3\,\tilde{\Lambda}\,\vtheta\,\vtheta'_L + \tilde{\Lambda}^L\,\vtheta'^R\,\vtheta''_{RL}.
\end{align}
Finally, after varying the remaining pieces in \eqref{RelLastInvariant}, which can be directly computed, one gets the variation of the last invariant:
\begin{align}
\delta_L\,[\tfrac{r^{-3}}{2}\,\vtheta''_{MP}\,\vtheta''^M{}_Q\,M^{PQ}] &\simeq 
r^{-3}\,\tilde{\Lambda}^N\,(3\,D^{(-1/2)}_N\,\vtheta
-M^{MP}\,D^{(-1/2)}_M\,\vtheta''_{NP}\nn\\
& -3\,f_{NPQ}\,M^{MQ} D^{(-1/2)}_M\,\vtheta'_P 
+\tfrac{1}{2}\,\vtheta\,\vtheta'_N\nn\\
& -\tfrac{1}{2}\,\vtheta'^M\,\vtheta''_{MN}
+M^{MP}\,\vtheta'_P\,\vtheta''_{MN}).
\end{align}
The final result for the variation of the ansatz of potential, up to total derivatives, is
\begin{align}
\delta_L\,V \simeq &\,-r^{-3}\,[(\tfrac{\alpha}{4} + \beta - 49\,\gamma + \tfrac{\xi}{2})\,\vtheta\,\vtheta'_M\,\tilde{\Lambda}^M
+(\beta + 21\,\gamma - \tfrac{\xi}{2})\,\vtheta''^{MN}\,\vtheta'_M\,\tilde{\Lambda}_N
\nn\\
& + (\xi-14\,\zeta)\,M^{MN}\,\vtheta'_M\,\vtheta''_{NP}\,\tilde{\Lambda}^P
-(\tfrac{\alpha}{4}+8\,\beta+98\,\gamma-3\,\xi)\,\tilde{\Lambda}^M\,D^{(-1/2)}_M\,\vtheta 
\nn\\
& + (14\,\zeta-\xi)\,M^{MN}\,\tilde{\Lambda}^P\,D^{(-1/2)}_N\,\vtheta''_{MP}
-(\vepsilon+3\,\xi)\,f^{MNP}\,M^Q{}_P\,D^{(-1/2)}_Q\,\vtheta'_N\,\tilde{\Lambda}_M],
\end{align}
which vanishes if and only if
\begin{equation}
\beta = -\tfrac{3}{40}\,\alpha - \tfrac{1}{15}\,\vepsilon, \quad
\gamma = \tfrac{1}{280}\,\alpha - \tfrac{1}{210}\,\vepsilon, \quad
\zeta = -\tfrac{1}{42}\,\vepsilon, \quad
\xi = - \tfrac{1}{3}\,\vepsilon.
\end{equation}
Therefore, redefining suitably $\alpha$ and $\vepsilon$, the most general potential reads
\begin{align}
V &= \alpha\,r^{-3}\,(\vtheta''_{MN}\,\vtheta''^{MN} - 21\,\vtheta'_M\,\vtheta'^M + 280\,\vtheta^2) \nn\\
& + \vepsilon\,r^{-3}\,(-32\,\vtheta^2 + \tfrac{5}{2}\,\vtheta'_M\,\vtheta'^M -\tfrac{3}{28}\,\vtheta''_{MN}\,\vtheta''^{MN} - 
\tfrac{3}{2}\,\vtheta'_M\,\vtheta'_N\,M^{MN} \nn\\
& + \tfrac{1}{28}\,\vtheta''_{MP}\,\vtheta''_{NQ}\,M^{MN}\,M^{PQ} + \tfrac{1}{2}\,\vtheta''_{MP}\,\vtheta''^M{}_Q\,M^{PQ}).\label{DoublePot}
\end{align}

So far, we have not used the section constraint. Now, observe that the combination in the first line is actually vanishing on section \cite{Inverso:2024xok}\footnote{An equivalent identity was also found in \cite{Eloy:2023zzh} ($\text{Eq.}$ 4.10).}
\begin{equation}\label{RelOnSec}
\vtheta''_{MN}\,\vtheta''^{MN} - 21\,\vtheta'_M\,\vtheta'^M + 280\,\vtheta^2 \approx 0.
\end{equation}
To see why, consider \eqref{Inv1}--\eqref{Inv3}, which imply that
\begin{align}
&a_1\,\vtheta^2 + a_2\,\vtheta'_M\,\vtheta'^M + a_3\,\vtheta''_{MN}\,\vtheta''^{MN} \nn\\
\approx\,& (\tfrac{31}{2}\,a_1\,P^{(1)} -60\,a_2\,P^{(248)} +196\,a_3\,P^{(3875)})^{MNPQ}\,W''_{MN}\,W''_{PQ}. \label{IdentityOnSection0}
\end{align}
Indeed, the section constraint ensures that $W'_M\,W'^M \approx 0 \approx \vtheta'_M\,W'^M$. Now, using \eqref{SecCon}, we are free to add any combination of ${P^{(n)}}^{MPNQ}\,W''_{MN}\,W''_{PQ}$, $n=1,248,3875$, since the first index in $W''_{MN}$ is on section, and try to find a choice of the constants which make vanishing the whole combination. Consider
\begin{equation}
(\tfrac{31}{2}\,a_1\,P^{(1)} -60\,a_2\,P^{(248)} + 196\,a_3\,P^{(3875)})^{MNPQ} + (b_1\,P^{(1)} +\,b_2\,P^{(248)} + b_3\,P^{(3875)})^{MPNQ}.
\end{equation}
Replacing \eqref{P1}--\eqref{P3875}, the combination above vanishes if and only if (up to an overall constant: fix, say, $a_3 = 1$)
\begin{equation}
a_1 = 280, \;
a_2 = -21, \;
a_3 = 1, \;
b_1 = -4340, \;
b_2 = -1260, \;
b_3 = -196,
\end{equation}
Replacing in \eqref{IdentityOnSection0},
\begin{align}
& 280\,\vtheta^2 -21\,\vtheta'_M\,\vtheta'^M + \vtheta''_{MN}\,\vtheta''^{MN} \nn\\
\approx\,& (\tfrac{31}{2}\,280\,P^{(1)} -60\cdot 21\,P^{(248)} +196\,P^{(3875)})^{MNPQ}\,W''_{MN}\,W''_{PQ} \nn\\
=&\, (4340\,P^{(1)} +1260\,P^{(248)} +196\,P^{(3875)})^{MPNQ}\,\,W''_{MN}\,W''_{PQ} \approx 0,
\end{align}
which is \eqref{RelOnSec}.

Therefore, the non-trivial solution in \eqref{DoublePot} is unique and equal to 
\begin{align}
r^3\,V &= -32\,\vtheta^2 + \tfrac{5}{2}\,\vtheta'_M\,\vtheta'^M -\tfrac{3}{28}\,\vtheta''_{MN}\,\vtheta''^{MN} - 
\tfrac{3}{2}\,\vtheta'_M\,\vtheta'_N\,M^{MN} \nn\\
& + \tfrac{1}{28}\,\vtheta''_{MP}\,\vtheta''_{NQ}\,M^{MN}\,M^{PQ} + \tfrac{1}{2}\,\vtheta''_{MP}\,\vtheta''^M{}_Q\,M^{PQ}.\label{PotTele}
\end{align}
This is precisely the potential in Eq. 3.71 of \cite{Hohm:2014fxa}, up to a total derivative, as it is checked in the next section.

\section{Comparison with the metric formulation}\label{Sec4}

Let us compare the potential of $n=8$ exceptional field theory in teleparallel formulation \eqref{PotTele} with the potential in ``metric" formulation, found in \cite{Hohm:2014fxa} and given by
\begin{align}
r^3\,\tilde{V} &= \tfrac{1}{4}\,j_M{}^R\,j_N{}^S\,(M^{MN}\,\eta_{RS} - 2\,M^{KL}\,f_{RL}{}^N\,f_{SK}{}^M + 2\,\delta^N_R\,\delta^M_S) \nn\\
& -\tfrac{1}{2}\,g^{-1}\,\de_M\,g\,M^{MN}\,f_{NK}{}^P\,j_P{}^K \nn\\
& + \tfrac{1}{4}\,M^{MN}\,g^{-1}\,\de_M\,g\,g^{-1}\,\de_N\,g - \tfrac{1}{4}\,M^{MN}\,\de_M\,g^{\mu\nu}\,\de_N\,g_{\mu\nu},\label{PotHS}
\end{align}
where $g_{\mu\nu}$ is the external metric, and the current $j_{MN}$ is defined as
\begin{equation}
j_{MN} := -\tfrac{1}{60}\,f_N{}^S{}_R\,M^{RQ}\,\de_M\,M_{QS}.
\end{equation}
Using
\begin{equation}
0 = D^{(-1/2)}_M\,M_{NP} = \de_M\,M_{NP} - W_M{}^L{}_{(N}\,M_{P)L},
\end{equation}
the current can be rewritten as
\begin{equation}
j_{MN} = (M_{NP}+\eta_{NP})\,\tfrac{1}{60}\,f^P{}_L{}^R\,W_M{}^L{}_R = (M_{NP}+\eta_{NP})\,W''_M{}^P.
\end{equation}
Let us first assume the Scherk-Schwarz ansatz for the metric $g_{\mu\nu}$,
\begin{equation}
g_{\mu\nu}(x,y) = r^{-2}(y)\,g_{\mu\nu}(x),
\end{equation}
where $x$ are the external coordinates and $y$ the internal ones. At the end of this section we will also take into account the additional terms which are present when $g_{\mu\nu}(x)$ is allowed to depend on $y$. The Scherk-Schwarz ansatz implies that
\begin{equations}
g^{-1}\,\de_M\,g &= g^{\mu\nu}\,\de_M\,g_{\mu\nu} = -6\,\de_M\log{r} = -3\,W'_M,\\
\de_M\,g_{\mu\nu}\,\de_N\,g^{\mu\nu} &= -12\,\de_M\log{r}\de_N\log{r} = -3\,W'_M\,W'_N,
\end{equations}
Replacing in the potential \eqref{PotHS}, one finds
\begin{align}
r^3\,\tilde{V} &= \tfrac{1}{4}\,(M^{PQ}+\eta^{PQ})\,M^{MN}\,W''_{MP}\,W''_{NQ} \nn\\
& - \tfrac{1}{2}\,M^{RS}\,(f_R{}^{PN} + f_{RL}{}^N\,M^{PL})\,(f_S{}^{QM} + f_{SK}{}^M\,M^{QK})\,W''_{MP}\,W''_{NQ} \nn\\
& + \tfrac{1}{2}\,(M^{NP}+\eta^{NP}\,(M^{MQ}+\eta^{MQ})\,W''_{MP}\,W''_{NQ} \nn\\
& + \tfrac{3}{2}\,(M^{MQ}\,f_Q{}^{PN} + M^{NQ}\,f_Q{}^{PM})\,W'_M\,W''_{NP} - \tfrac{3}{2}\,M^{MN}\,W'_M\,W'_N.\label{PotHS1}
\end{align}
Using the Jacobi identity \eqref{DecJacobi} and the relations \eqref{RelM}--\eqref{RelMlast}, and observing that
\begin{align}
& r^{-3}\,(\tfrac{3}{2}\,M^{RM}\,f^{NP}{}_R\,W'_{[M}\,W''_{N]P} + M^{RS}\,f^{PQ}{}_R\,W''_{PQ}\,f^{MN}{}_S\,W''_{MN} \nn\\
&\qquad  - M^{RS}\,f^{NP}{}_R\,f^{MQ}{}_S\,W''_{MP}\,W''_{NQ}) = \de_M\,(r^{-3}\,\delta^{AB}\,U^{[M}{}_A\,\de_N\,U^{N]}{}_B) \nn\\
&\;\qquad\qquad\qquad\qquad\qquad\qquad\qquad\qquad = \de_M\,(r^{-3}\,M^{R[M}\,f^{N ]P}{}_R\,W''_{NP}),
\end{align}
one can rewrite the potential \eqref{PotHS1} in the following way:
\begin{align}
\tilde{V} = V_1 + V_2 + V_3 + V_4 + \de_M\,(r^{-3}\,M^{R[M}\,f^{N]P}{}_R\,W''_{NP}),\label{V0}
\end{align}
where 
\begin{align}
r^3\,V_1 &= \tfrac{1}{2}\,[(M^{MQ}\,M^{NP} + M^{MN}\,M^{PQ}) \nn\\
& - \tfrac{1}{2}\,(f^{PQ}{}_{RS}\,M^{MR}\,M^{NS} + f^{MNR}\,f^{PQ}{}_S\,M^{RS})]\,W''_{MN}\,W''_{PQ},\\
r^3\,V_2 &= M^{MP}\,W''_{MN}\,W''^N{}_P + \tfrac{1}{2}\,M^{MP}\,W''_{MN}\,W''_P{}^N \,\nn\\
& - f^{PNS}\,f^Q{}_{RS}\,M^{MR}\,W''_{MN}\,W''_{PQ} \nn\\
& -\tfrac{1}{2}\,(f^{MQ}{}_R\,f^{NP}{}_{SG} - f^{MN}{}_R\,f^{PQ}{}_S)\,M^{RS}\,W''_{MN}\,W''_{PQ},\\
r^3\,V_3 &= -3\,W''_{MN}\,W''^{MN} - \tfrac{7}{2}\,W''_M{}^M\,W''_N{}^N,\\
r^3\,V_4 &= -\tfrac{3}{2}\,M^{MN}\,W'_M\,W'_N + \tfrac{7}{2}\,W''_{MN}\,W''^{NM} + \tfrac{7}{2}\,W''_M{}^M\,W''_N{}^N \nn\\
& - 3\,f^{NP}{}_Q\,M^{MQ}\,W'_M\,W''_{NP} - \tfrac{3}{2}\,f^{MN}{}_R\,f^{PQ}{}_S\,M^{RS}\,W''_{MN}\,W''_{PQ}.
\end{align}
Notice that the first piece can be written in the following way
\begin{equation}
r^3\,V_1 = \tfrac{1}{2}\,M^{MN}\,M^{PQ}\,W''_{MP}\,(W_{(NP)}-f^S{}_{NL}\,f^R{}_Q{}^L\,W''_{RS}).\label{V1}
\end{equation}
Adding terms which vanish on section, one can rewrite the second piece as
\begin{align}
r^3\,V_2 &\approx r^3\,V_2 +
\tfrac{1}{2}\,M^{MN}\,W''_{PM}\,W''^P{}_N
+M^{NR}\,f^Q{}_{RS}\,f^{MPS}\,W''_{MN}\,W''_{PQ} \nn\\
& +\tfrac{1}{2}\,M^{RS}\,f_R{}^{LN}\,f_S{}^{IQ}\,(f^M{}_{LK}\,f^P{}_I{}^K\,W''_{MN}\,W''_{PQ} - W''_{LN}\,W''_{IQ} -W''_{IN}\,W''_{LQ}) \nn\\
& =\tfrac{1}{2}\,M^{MN}\,\eta^{PQ}\,(W''_{(MP)} - f^S{}_{ML}\,f^R{}_N{}^L\,W''_{RS})\,(W''_{(NQ)}-f^S{}_{NL}\,f^R{}_Q{}^L\,W''_{RS}).\label{V2}
\end{align}
Similarly for the third piece
\begin{align}
r^3\,V_3 &\approx r^3\,V_3 - \tfrac{3}{2}\,W''_{MN}\,W''^{MN} \nn\\
& - \tfrac{3}{2}\,(f^{PMQ}\,f^{RN}{}_Q\,W''_{PM}\,W''_{RN} - W''_M{}^M\,W''_N{}^N - W''_M{}^N\,W''_N{}^M) \nn\\
& = -\tfrac{3}{2}\,\eta^{MN}\,\eta^{PQ}\,W''_{MP}\,(W''_{(NQ)}-f^S{}_{NL}\,f^R{}_Q{}^L\,W''_{RS})-2\,W''_M{}^M\,W''_N{}^N,\label{V3}
\end{align}
and for the fourth one
\begin{align}
r^3\,V_4 & \approx r^3\,V_4 + \tfrac{7}{2}\,(f^{PMQ}\,f^{RN}{}_Q\,W''_{PM}\,W''_{RN} - W''_M{}^M\,W''_N{}^N - W''_M{}^N\,W''_N{}^M) \nn\\
& = \tfrac{7}{2}\,f^{MN}{}_R\,f^{PQR}\,W''_{MN}\,W''_{PQ} - \tfrac{3}{2}\,M^{RS}\,f^{MN}{}_R\,f^{PQ}{}_S\,W''_{MN}\,W''_{PQ} \nn\\
& -3\,W'_M\,W''_{NP}\,f^{NP}{}_Q\,M^{QM} - \tfrac{3}{2}\,M^{MN}\,W'_M\,W'_N.\label{V4}
\end{align}
In this way, it is easy to see that each piece can be written in terms of the components of the torsion. Indeed, the following identities hold
\begin{equation}
W''^{(27000)}_{MN}\,\vtheta''^{MN} = 0, \quad
W''^{(27000)}_{MN}\,\vtheta''_{PQ}\,M^{MP}\,M^{NQ} = 0, \quad
W''^{(30380)}_{MN}\,\vtheta''_{PQ}\,M^{MP}\,M^{NQ} = 0,
\end{equation}
since $W''^{(27000)}_{MN}$, $W''^{(30380}_{MN}$ and $\vtheta''_{MN}$ (the last one in the representation $\bm{3875}$) are orthogonal irreducible components of $W''_{MN}$. They imply that
\begin{equation}
\vtheta''_{MN}\,\vtheta''^{MN} = 14\,W''_{MN}\,\vtheta''^{MN}, \quad
\vtheta''_{MN}\,\vtheta''_{PQ}\,M^{MP}\,M^{NQ} = 14\,W''_{MN}\,\vtheta''_{PQ}\,M^{MP}\,M^{NQ}.
\end{equation}
Moreover, using the Jacobi identity \eqref{DecJacobi}, one sees that the following combination can be written in terms of the components of the torsion
\begin{equation}
W''_{MP} + W''_{PM} - f^S{}_{ML}\,f^R{}_P{}^L\,W''_{RS} = \vtheta''_{MP} + \eta_{MP}\,\vtheta + \tfrac{1}{2}\,f_{MP}{}^L\,(\vtheta'_L - W'_L).
\end{equation}
Using these observations, all the pieces $V_i$, $i=1,2,3,4$, can be written in terms of the components of the torsion. Namely, for the first piece in \eqref{V1}, 
\begin{equation}
r^3\,V_1 = \tfrac{1}{28}\,M^{MN}\,M^{PQ}\,\vtheta''_{MP}\,\vtheta''_{NP} + 2\,\vtheta^2 - \tfrac{1}{4}\,M^{MN}\,(\vtheta'_M-W'_M)\,(\vtheta'_N-W'_N).
\end{equation}
For the second piece in \eqref{V2},
\begin{align}
r^3\,V_2 &\approx \tfrac{1}{2}\,M^{MN}\,\vtheta''_{MP}\,\vtheta''_N{}^P - \tfrac{1}{4}\,M^{MN}\,(\vtheta'_M-W'_M)\,(\vtheta'_N-W'_N) \nn\\
& - \tfrac{1}{4}\,(\vtheta'_M-W'_M)\,(\vtheta'^M-W'^M) + 4\,\vtheta^2 \nn\\
& \approx \tfrac{1}{2}\,M^{MN}\,\vtheta''_{MP}\,\vtheta''_N{}^P - \tfrac{1}{4}\,M^{MN}\,(\vtheta'_M-W'_M)\,(\vtheta'_N-W'_N) \,\nn\\
& - \tfrac{1}{4}\,\vtheta'_M\,\vtheta'^M + 4\,\vtheta^2.
\end{align}
For the third piece in \eqref{V3},
\begin{align}
r^3\,V_3 &\approx -\tfrac{3}{28}\,\vtheta''_{MN}\,\vtheta''^{MN} - \tfrac{3}{4}\,(\vtheta'_M-W'_M)\,(\vtheta'^M-W'^M) - 38\,\vtheta^2 \nn\\
& \approx -\tfrac{3}{28}\,\vtheta''_{MN}\,\vtheta''^{MN} - \tfrac{3}{4}\,\vtheta'_M\,\vtheta'^M - 38\,\vtheta^2.
\end{align}
Finally, for the fourth piece in \eqref{V4},
\begin{align}
r^3\,V_4 &\approx \tfrac{7}{2}\,(\vtheta'_M-W'_M)\,(\vtheta'^M-W'^M) - \tfrac{3}{2}\,M^{MN}\,\vtheta'_M\,\vtheta'_N \nn\\
& \approx \tfrac{7}{2}\,\vtheta'_M\,\vtheta'^M - \tfrac{3}{2}\,M^{MN}\,\vtheta'_M\,\vtheta'_N.
\end{align}
Replacing in \eqref{V0},
\begin{align}
r^3\,\tilde{V} &\approx -32\,\vtheta^2 + \tfrac{5}{2}\,\vtheta'_M\,\vtheta'^M - \tfrac{3}{28}\,\vtheta_{MN}''\,\vtheta''^{MN} - \tfrac{3}{2}\,M^{MN}\,\vtheta'_M\,\vtheta'_N \nn\\
& + \tfrac{1}{28}\,M^{MN}\,M^{PQ}\,\vtheta''_{MP}\,\vtheta''_{NQ} + \tfrac{1}{2}\,M^{PQ}\,\vtheta''_{MP}\,\vtheta''^M{}_Q \nn\\
& + \de_M\,(r^{-3}\,M^{R[M}\,f^{N]P}{}_R\,W''_{NP}),
\end{align}
which is the potential in teleparallel formulation \eqref{PotTele}, up to a total derivative:
\begin{equation}\label{FinalComp}
\tilde{V} \approx V + r^{-3}\,\de_M\,(r^{-3}\,M^{R[M}\,f^{N]P}{}_R\,W''_{NP}).
\end{equation}

We conclude this section by performing the comparison without using the Scherk-Schwarz ansatz. In this case, the potential written in terms of the quadratic scalar contractions of the torsion has to be supplemented by a kinetic piece for the external metric $g_{\mu\nu}$. Indeed, the external metric is treated as a singlet in exceptional field theory. From the perspective of the internal space, this is analogous to coupling the Hilbert–Einstein action for the internal space to scalar fields.

Define the breaking of the covariance as the difference between the variation under (generalised) change of coordinates $\delta_\xi$ and the generalised Lie derivative:
\begin{equation}
\Delta_\xi = \delta_\xi - L_\xi.
\end{equation}
The expression for the potential found in \cite{Hohm:2014fxa} is fixed by requiring that
\begin{equation}
\Delta_\xi\,\tilde{V} = 0.
\end{equation}
In particular, it is shown that the $\Delta_\xi$ variation of the terms in the potential depending on the external metric cancels precisely the variation of the terms quadratic in the current, up to total derivatives. The breaking of the covariance of the external metric is due to its determinant. Indeed, the previous computation has shown that the $r$-part in the Scherk-Schwarz ansatz is sufficient to reproduce the whole potential, up to a total derivative. Therefore, in order to extend the previous result without assuming the  Scherk-Schwarz ansatz, it is convenient to define a unimodular metric $\tilde{g}_{\mu\nu}$, depending in general both on the internal and on the external coordinates, according to 
\begin{equation}
g_{\mu\nu}(x,y) =: r^{-2}(x,y)\,\tilde{g}_{\mu\nu}(x,y), \;\;\text{with}\;\;\text{det}\,\tilde{g}_{\mu\nu}=1.\label{gext}
\end{equation} 
This implies that the determinant of the external metric is $g = r^{-6}$. Allowing a dependence on the external coordinates in the determinant factor does not modify the previous computation, since only the (extended) internal derivatives $\de_M$ are involved in the potential. 

The breaking of the covariance of $W'_M$ in \eqref{SpoilW1} implies that 
\begin{equation}
\Delta_\xi(g^{\mu\nu}\,\de_M\,g_{\mu\nu}) = 
\Delta_\xi(g^{-1}\,\de_M\,g) = \Delta_\xi\,\de_M\log{g} = 6\,\de_M\,\de_N\,\xi^N,
\end{equation}
which fixes the breaking of the covariance of the partial derivative of the external metric:
\begin{equation}
\Delta_\xi\,\de_M\,g_{\mu\nu} = 2\,g_{\mu\nu}\,\de_M\,\de_N\,\xi^N,
\end{equation}
using the fact that $g_{\mu\nu}$ is a three-dimensional metric. On the other hand, 
\begin{equation}\label{Covguni}
\Delta_\xi\,\de_M\,\tilde{g}_{\mu\nu} = 0,
\end{equation}
because of the unimodular condition $\tilde{g}^{\mu\nu}\,\de_M\,\tilde{g}_{\mu\nu} = 0$.

Consider now the terms in \eqref{PotHS} depending on the external metric, and replace \eqref{gext}. There is only one additional term with respect to \eqref{PotHS1}, coming from the last $M^{MN}\,\de_M\,g_{\mu\nu}\,\de_N\,g^{\mu\nu}$, because the other terms depending on the external metric involve only the determinant. This term is of the same form, with the unimodular metric in place of the full external metric. Therefore, the result \eqref{FinalComp} is extended without assuming the Scherk-Scwarz ansatz by
\begin{equation}
\tilde{V} \approx V - \tfrac{1}{4}\,r^{-3}\,M^{MN}\,\de_M\,\tilde{g}_{\mu\nu}\,\de_N\,\tilde{g}^{\mu\nu} + r^{-3}\,\de_M\,(r^{-3}\,M^{R[M}\,f^{N]P}{}_R\,W''_{NP}),
\end{equation}
where now, with an abuse of notation, we mean that the fields depend both on the internal and on the external coordinates. The additional term is covariant, because of \eqref{Covguni}.

In summary, we have learnt that the potential found in \cite{Hohm:2014fxa} is equivalent, up to a total derivative, to a ``genuine" potential, which reads as a combination of quadratic scalar invariants of the components of the intrinsic torsion, plus a ``kinetic term" for the unimodular metric. (Of course, the latter necessarily involves the generalised internal metric, since a term such as $\de_M\,\tilde{g}_{\mu\nu}\,\de^M\,\tilde{g}^{\mu\nu}$ is obviously vanishing on section). The advantage of the teleparallel formulation is that the covariance of the potential is now manifest, since both the derivatives of the unimodular metric and the components of the intrinsic torsion are covariant on section (see Appendix \ref{AppDiff} for details).

\section{Comparison with the supergravity potential}\label{Sec5}

In this section, the potential 
in teleparallel formulation \eqref{PotTele} is compared with the supergravity potential in three-dimensional gauged supergravities, found in \cite{Nicolai:2000sc, Nicolai:2001sv}. This was first done in \cite{GuillaumeNotes}.\footnote{The author thanks Guillaume Bossard for making his unpublished notes available.}

The adjoint/fundamental representation of the algebra of $E_{8(8)}$ decomposes with respect to its maximal
compact subgroup $\text{SO}(16)$ according to
\begin{equation}
\bm{248} \rightarrow \bm{128} \oplus \bm{120},
\end{equation}
where $\bm{128}$ is the chiral spinor representation and $\bm{120}$ is the adjoint representation of $\text{SO}(16)$
($\frac{16(16-1)}{2} =120$, $2^{16/2-1}=128$). Therefore a tensor $T_M$ decomposes in $(T_\alpha,T_{ij})$, where
$\alpha = 1,\dots,128$ is the undotted spinor index (similarly for the dotted index $\dot{\alpha}$ in $\overline{\bm{128}}$), and $i=1,\dots, 16$ is the fundamental index, and $T_{ij}=-T_{ji}$. Therefore, for the components $\vtheta''_{MN}$ of the generalised torsion, one has
\begin{equation}
\vtheta''_{MN} \rightarrow (\vtheta''_{\alpha\beta}, \vtheta''_{ij,\alpha}, \vtheta''_{ij,kl}),
\end{equation}
where
\begin{equation}
\vtheta''_{\alpha\beta} = \vtheta''_{\beta\alpha}, \quad
\vtheta''_{ij,\alpha} = -\vtheta''_{ji,\alpha}, \quad
\vtheta_{ij,kl} = -\vtheta''_{ji,kl} = -\vtheta''_{ij,lk} = \vtheta''_{kl,ij}, \quad 
\vtheta''_\alpha{}^\alpha - \vtheta''_{ij}{}^{ij} = 0,
\end{equation}
where the last condition follows from $\vtheta''_M{}^M = 0$. The decomposition of the representation $\bm{3875}$, in which $\vtheta''_{MN}$ transforms, is given by
\begin{equation}
\bm{3875} \rightarrow \bm{135} \oplus \bm{1820} \oplus \overline{\bm{1920}},
\end{equation}
where $\bm{135}$ is the rank-two symmetric traceless representation ($135 = \frac{16(16+1)}{2}-1$), 
\begin{equation}
\Xi_{ij} = \Xi_{ji}, \quad \Xi_i{}^i = 0,
\end{equation}
$\bm{1820}$ is the rank-four totally antisymmetric representation ($1820 = \tfrac{16(16-1)(16-2)(16-3)}{24}$),
\begin{equation}
\Xi_{ijkl} = \tfrac{1}{24}\,\Xi_{[ijkl]},
\end{equation}
and $\overline{\bm{1920}}$ is the dual traceless vector spinor representation ($1920 = 16\cdot 128 - 128$),
\begin{equation}
\Xi_{i,\dot{\alpha}}\;\;\text{such that}\;\;(\Gamma_i)_{\alpha\dot{\alpha}}\,\Xi^{i,\dot{\alpha}} = 0,
\end{equation}
where $(\Gamma_i)_{\alpha\dot{\alpha}}$ are the components of the 16 $128 \times 128$ gamma matrices (the 16-dimensional analogue of the Pauli matrices). Therefore, the irreducible components of $\vtheta''_{MN}$ are $(\Xi_{ij},\Xi_{i,\alpha},\Xi_{ijkl})$. The relation between the $\vtheta''$'s and the $\Xi$'s can be obtained in the following way: $\vtheta''_{ij,\alpha}$ is the dual of $\Xi_{i,\dot{\alpha}}$, that is, $\vtheta''_{ij,\alpha} = \gamma\,(\Gamma_{[i})_{\alpha\dot{\alpha}}\,\Xi_{j]}{}^{\dot{\alpha}}$; $\vtheta''_{\alpha\beta}$ is necessarily proportional to $(\Gamma^{ijkl})_{\alpha\beta}\,\Xi_{ijkl}$, where $\Gamma^{ijkl}$ is the usual antisymmetrised product of four $\Gamma^i$ matrices, that is, $\vtheta''_{\alpha\beta} = \beta\,(\Gamma^{ijkl})_{\alpha\beta}\,\Xi_{ijkl}$; finally, $\vtheta''_{ij,kl}$ is necessarily  a combination of $\Xi_{ijkl}$ and of $\Xi_{ij}$, that is, $\vtheta''_{ijkl} = \Xi_{ijkl} + \alpha\,\delta_{[i[k}\,\Xi_{l]j]}$. (Notice that, in this way, $\vtheta''_\alpha{}^\alpha = 0 =\vtheta''_{ij}{}^{ij}$, as expected.) Choosing the following normalisation,
\begin{equation}
\vtheta''_{ij,\alpha} = -\tfrac{1}{4}\,\tfrac{1}{14}\,(\Gamma_{[i})_{\alpha\dot{\alpha}}\,\Xi_{j]}{}^{\dot{\alpha}}, \quad
\vtheta''_{\alpha\beta} = \tfrac{1}{4}\,\tfrac{1}{96}\,(\Gamma^{ijkl})_{\alpha\beta}\,\Xi_{ijkl},\quad
\vtheta''_{ijkl} = \tfrac{1}{4}(\,\Xi_{ijkl} + \delta_{[i[k}\,\Xi_{l]j]}),
\end{equation}
one can compute
\begin{equations}
& \vtheta''_{\alpha\beta}\,\vtheta''^{\alpha\beta} = \tfrac{1}{48}\,\Xi_{ijkl}\,\Xi^{ijkl},\\
& \vtheta''_{ij,\alpha}\,\vtheta''^{ij,\alpha} = \tfrac{1}{112}\,\Xi_{i,\dot\alpha}\,\Xi^{i,\dot\alpha}, \\
& \vtheta''_{ij,kl}\,\vtheta''^{ij,kl} = \tfrac{1}{16}\,\Xi_{ijkl}\,\Xi^{ijkl} + \tfrac{7}{2}\,\Xi_{ij}\,\Xi^{ij},
\end{equations}
where one has to use $\text{Tr}\,(\Gamma^{ijkl}\,\Gamma_{abcd}) = 128\,\delta^i_{[a}\,\delta^j_b\,\delta^k_c\,\delta^l_{d]}$. 

Observe that, setting $\vtheta' \rightarrow 0$ in the potential in teleparallel formulation \eqref{PotTele},
\begin{align}
V_{|_{\vtheta' \rightarrow 0}} =&\,
-32\,\vtheta^2 + \tfrac{1}{28}\,(M^{MN}\,M^{PQ}\,\vtheta''_{MP}\,\vtheta''_{NQ} + 14\,M^{MN}\,\vtheta''_{MP}\,\vtheta''_N{}^P-3\,\vtheta''_{MN}\,\vtheta''^{MN}) \nn\\
=&\,-32\,\vtheta^2+\tfrac{1}{28}\,[\text{Tr}\,(\vtheta''\,\delta\,\vtheta''\,\delta) + 14\,\text{Tr}\,(\vtheta''\,\delta\,\vtheta''\,\eta) -3\,\text{Tr}\,(\vtheta''\,\eta\,\vtheta''\,\eta)] \nn\\
=&\,-32\,\vtheta^2+\tfrac{1}{28}\,[(\vtheta''_{\alpha\beta}\,\vtheta''^{\alpha\beta} + 2\,\tfrac{1}{2}\,\vtheta''_{ij,\alpha}\,\vtheta''^{ij,\alpha} + \tfrac{1}{4}\,\vtheta''_{ij,kl}\,\vtheta''^{ij,kl}) \nn\\
+&\, 14\,(\vtheta''_{\alpha\beta}\,\vtheta''^{\alpha\beta} - \tfrac{1}{2}\,\vtheta''_{ij,kl}\,\vtheta''^{ij,kl})
-3\,(\vtheta''_{\alpha\beta}\,\vtheta''^{\alpha\beta} - 2\,\tfrac{1}{2}\,\vtheta''_{ij,\alpha}\,\vtheta''^{ij,\alpha} + \tfrac{1}{4}\,\vtheta''_{ij,kl}\,\vtheta''^{ij,kl})] \nn\\
=&\,-32\,\vtheta^2+\tfrac{1}{28}\,(12\,\vtheta''_{\alpha\beta}\,\vtheta''^{\alpha\beta} + 4\,\vtheta''_{ij,\alpha}\,\vtheta''^{ij,\alpha} -4\,\vtheta''_{ij,kl}\,\vtheta''^{ij,kl}) \nn\\
=&\, \tfrac{1}{784}\,\Xi_{i,\dot\alpha}\,\Xi^{i,\dot\alpha}-\tfrac{1}{2}\,\Xi_{ij}\,\Xi^{ij}.
\end{align}
Finally, defining\footnote{In \cite{Nicolai:2000sc, Nicolai:2001sv} (see also \cite{deWit:2003fgi}) the embedding tensor $\Theta_{MN}$, which takes values in the representations
$\bm{1} \oplus \bm{3875}$, is decomposed in $\text{SO}(16)$ irreducible components according to
\begin{equations}
\Theta_{\alpha,\beta} &= \theta\,\delta_{\alpha\beta} + \tfrac{1}{96}\,(\Gamma^{ijkl})_{\alpha\beta}\,\Xi_{ijkl},\label{Theta1}\\
\Theta_{ij,\alpha} &= -\tfrac{1}{14}\,(\Gamma_{[i})_{\alpha\dot\alpha}\,\Xi_{j]}{}^{\dot\alpha},\label{Theta2}\\
\Theta_{ij,kl} &= -2\,\theta\,\delta_{ij} + \delta_{[i[k}\,\Xi_{l]j]} + \Xi_{ijkl},\label{Theta3}
\end{equations}
where $\theta$ is the singlet, and $\Xi_{ij}$, $\Xi_{ijkl}$, and $\Xi_{i,\dot\alpha}$ are in $\bm{135}$, $\bm{1820}$, and $\bm{1920}$ respectively. The tensors $A_1$ and $A_2$ are given in terms of the $\Theta$'s and of $\theta$:
\begin{equation}
(A_1)_{ij} = \tfrac{8}{7}\,\theta\,\delta_{ij} + \tfrac{1}{7}\,\Theta_{ik,j}{}^k, \quad
(A_2)_{i,\dot\alpha} = -\tfrac{1}{7}\,(\Gamma^j)_{\alpha\dot\beta}\,\Theta_{ij}{}^{\alpha},
\end{equation}
which, using \eqref{Theta1}--\eqref{Theta3} and setting $\theta = 4\,\vtheta$, are written as in \eqref{Atensors}.}
\begin{equation}\label{Atensors}
(A_1)_{ij} = 2\,\Xi_{ij} - 4\,\vtheta\,\delta_{ij}, \quad
(A_2)_{i,\dot\alpha} = -\tfrac{1}{7}\,\Xi_{i,\dot\alpha}.
\end{equation}
one gets 
\begin{equation}
V_{|_{\vtheta' \rightarrow 0}}  = -\tfrac{1}{8}\,[(A_1)^2 - \tfrac{1}{2}\,(A_2)^2],
\end{equation}
which is the potential found in \cite{Nicolai:2000sc, Nicolai:2001sv}.

\section{Conclusions and discussion}\label{Sec6}

One may ask if it is possible to select the potential \eqref{PotTele} from the variation of the most general ansatz without using the condition \eqref{RelOnSec}, which, as mentioned before, is not satisfied in general by the embedding tensor, when the components of the intrinsic torsion are identified with those of the embedding tensor.

A path that can be in principle explored is to assume that the frame does \emph{not} satisfy the section constraint \eqref{SectionConstraintY}. This means that, when a derivative acts on the frame, the section constraint is not satisfied:
\begin{equation}
Y^{MP}{}_{QN}\,\de_M E_R{}^A\,U_P \neq 0, \quad
Y^{MP}{}_{QN}\,U_M\,\de_P E_R{}^A \neq 0, 
\end{equation}
where $U_P$ is any partial derivative or an ancillary parameter.

Following the idea developed in \cite{Ciceri:2016dmd}, one may  parametrise the breaking of the section constraint by the frame by dressing (or twisting) it with an invertible matrix, such that the dressed frame does satisfy the constraint. Let $E^M{}_A$ be the frame which does not satisfy the section constraint. Thus, the first index of the Weitzenb\"ock connection $W_M{}^P{}_N$ defined by $E^M{}_A$, that is, the first index of its components $W'_M$ and $W''_{MN}$ is not on section. Denote the dressing matrix $C^M{}_{N'}$, so that
\begin{equation}
E^M{}_A = C^M{}_{N'}\,\tilde{E}^{N'}{}_A,
\end{equation}
where $\tilde{E}^M{}_A$ is the frame which does satisfy the section constraint. The prime in the second index of $C^M{}_{N'}$ and in the curved index of $\tilde{E}^{M'}{}_A$ is only for bookkeeping. 

It is possible to derive the relation between the components of the torsion associated to the frame $E^M{}_A$, of the one associated to the frame $\tilde{E}^{}_A$, and of the one associated to the (inverse) dressing matrix $C^M{}_{M'}$. Upon identifying the latter with the components of the embedding tensor, one assumes that they are constant and satisfy the quadratic constraint. 

Finally, one should repeat the computation of the Lorentz variation of the ansatz of the potential, considering that the Bianchi identities \eqref{Bianchi4}--\eqref{Bianchi3} are not guaranteed to be satisfied by the components of the torsion of the frame $E^M{}_A$. Then, one may look for additional constraints that allow one to select the potential compatible with the supergravity potential of three-dimensional gauged supergravity. 

Not only is this procedure somehow involved and \emph{ad hoc}, but it also suffers from a more serious problem: as shown in Appendix \ref{AppDiff}, the components of the torsion are genuine tensor (with respect to the generalised diffeomorphisms, that is, their infinitesimal transformations under generalised diffeomorphisms is generated by the generalised Lie derivative) \emph{only} on section. So, if the frame does not satisfy the section constraint, the components of the corresponding intrinsic torsion are not genuine tensors, and the ansatz for the potential is no longer automatically a scalar density under generalised diffeomorphism. Therefore, one must impose this as an additional requirement.

In this case, the teleparallel approach would lose its effectiveness: instead of starting from an ansatz that is manifestly invariant under diffeomorphisms (up to a total derivative) and imposing only invariance under local Lorentz transformations, one would also have to impose generalised diffeomorphism invariance. Moreover, the ansatz would also have to be supplemented in general by the most general combination of independent terms which are zero on section, such as $W'_M\,W'^M$ or $f^{MNP}\,W'_M\,W''_{NP}$.

Even taking all these subtleties and difficulties into account, we have not yet been able to find an effective way of selecting the desired potential \eqref{PotTele} without the additional condition \eqref{RelOnSec} implied by the section constraint.

\section*{Acknowledgements}

The author is deeply indebted to Guillaume Bossard, who suggested the problem and devoted much of his time to discussions and to providing suggestions, comments, and ideas. This work is supported by the Fondazione Angelo della Riccia grant 2026 and by DDFIP Essonne. 

\appendix

\section{Invariant tensors}\label{App1}

Denote by $f^{MNR}$ the structure constants of $E_{8(8)}$, and by $f^{MP}{}_{NQ}$ the symmetrised product of two structure constants (without numerical coefficients):
\begin{equation}
f^{MP}{}_{NQ} := f^{RM}{}_N\,f_R{}^P{}_Q + f^{RP}{}_N\,f_R{}^M{}_Q.
\end{equation}
The Jacobi identity 
\begin{equation}
f^{RM}{}_N\,f_R{}^{PQ} +
f^{RP}{}_N\,f_R{}^{QM} +
f^{RQ}{}_N\,f_R{}^{MP} = 0
\end{equation}
can be written as
\begin{equation}\label{DecJacobi}
f^{RM}{}_N\,f_R{}^P{}_Q = 
\tfrac{1}{2}\,f^{RMP}\,f_{RNQ} + \tfrac{1}{2}\,f^{MP}{}_{NQ}.
\end{equation}

The projectors on the first few irreducible representations are \cite{Koepsell:1999uj}
\begin{equations}
& {P^{(1)}}_{MN}{}^{PQ} = \tfrac{1}{248}\,\eta_{MN}\,\eta^{PQ}, \label{P1}\\
& {P^{(248)}}_{MN}{}^{PQ} = -\tfrac{1}{60}\,f_{MNR}\,f^{PQR}, \label{P248}\\
& {P^{(3875)}}_{MN}{}^{PQ} = \tfrac{1}{14}\,\delta_M^P\,\delta_N^Q + \tfrac{1}{14}\,\delta_M^Q\,\delta_N^P - \tfrac{1}{56}\,\eta_{MN}\,\eta^{PQ} - \tfrac{1}{28}\,f_{MN}{}^{PQ},\label{P3875}\\
& {P^{(27000)}}_{MN}{}^{PQ} = \tfrac{3}{7}\,\delta_M^P\,\delta_N^Q + \tfrac{3}{7}\,\delta_M^Q\,\delta_N^P + \tfrac{3}{217}\,\eta_{MN}\,\eta^{PQ} + \tfrac{1}{28}\,f_{MN}{}^{PQ}, \label{P27000}\\
& {P^{(30380)}}_{MN}{}^{PQ} = \tfrac{1}{2}\,\delta_M^P\,\delta_N^Q - \tfrac{1}{2}\,\delta_M^Q\,\delta_M^P + \tfrac{1}{60}\,f_{MNR}\,f^{PQR}.\label{P30380}
\end{equations}

Here is a collection of useful identities:\footnote{A possible parametrisation is $\eta_{MN} = \text{diag}\,(-,\dots,-,+,\dots,+)$ and $\delta_{AB} = \text{diag}\,(+,\dots,+,+,\dots, +)$, according to the decomposition of the adjoint/fundamental $\bf{248}$ of $E_{8(8)}$ in $\bf{120}\oplus\bf{128}$ of its maximal compact subgroup $\text{SO}(16)$.}
\begin{equations}
& \eta_{AB}\,\delta^{AB} = 8, \label{propDelta1}\\
& \eta_{AB}\,\eta^{AB} = \delta_{AB}\,\delta^{AB} = 248,\label{propDelta2}\\
& \delta^{AD}\,\delta^{BE}\,\delta^{CF}\,f_{ABC} = - f^{DEF},\label{propDelta3}\\
& f^{MRT}\,f^{NS}{}_T\,f_{PR}{}^U\,f_{QSU} = 12\,\delta^M_P\,\delta^N_Q + 12\,\delta^M_Q\,\delta^N_P + 12\,\eta^{MN}\,\eta_{PQ} \nn\\
& \qquad\qquad\qquad\qquad\qquad\quad - 5\,f^{MN}{}_{PQ} - 15\,f^{MNR}\,f_{PQR},\\
& f_{MNRS}\,f_{PQ}{}^{RS} = - 20\,f_{MNPQ} + 48\,(\eta_{MP}\,\eta_{NQ} + \eta_{MQ}\,\eta_{NP} + \eta_{MN}\,\eta_{PQ}),\\
& f_P{}^{QR}\,f_{MQNR} = -90\,f_{PMN},\\
& f^{ACEF}\,f^{BD}{}_{EF} =
48\,(\eta^{AC}\,\eta^{BD}
+\eta^{BC}\,\eta^{AD}
+\eta^{AB}\,\eta^{CD}) \nn\\
& \qquad\qquad\qquad\quad +10\,f^{ABCD} -30\,f^{EAB}\,f_E{}^{CD}.
\end{equations}
The following expression is a consequence of the section constraint \eqref{SectionConstraintY} and of the Jacobi identity \eqref{DecJacobi} (see Appendix A1 in \cite{Hohm:2014fxa}): 
\begin{equation}
(f^R{}_{UV}\,f^{UP}{}_K\,f^{VQ}{}_L +\delta^P_{[K}\,f^{QR}{}_{L]} + \eta^{RP}\,f^Q{}_{KL})\,(U_P\,V_Q+U_Q\,V_P) \approx 0.
\end{equation}
In particular, this implies
\begin{equation}\label{FsymSecCon}
\tfrac{1}{2}\,f_{MN}{}^{PQ}\,f^{KL}{}_Q\,(U_L\,V_P+ U_P\,V_L) \approx f^{KL}{}_{(M}\,(U_{N)}\,V_L + U_L\,V_{N)}).
\end{equation}

\section{Behaviour under diffeomorphisms of a connection}\label{AppDiff}

It is convenient to parametrise the $n=8$ generalised Lie derivative of a vector with density weight $\lambda = 1$ in the following way:
\begin{equation}
L_{\xi,\sigma}\,V^M = \xi^N\,\de_N\,V^M + Z^{MN}{}_{PQ}\,\de_N\,\xi^P\,V^Q + f^{MP}{}_N\,\sigma_P\,V^N,
\end{equation}
where 
\begin{equation}\label{ZE8}
Z^{MN}{}_{PQ} = -\delta^M_P\,\delta^N_Q + Y^{MN}{}_{PQ} = \delta^M_Q\,\delta^N_P - f^{RM}{}_Q\,f_R{}^N{}_P.
\end{equation}
The $E_{n(n)}$ exceptional geometry with $n\leqslant 7$ is obtained by setting $\sigma(\xi)\rightarrow 0$. Ordinary geometry is obtained by requiring also $Y^M \rightarrow x^\mu$, $Z^{MR}{}_{SN} \rightarrow \delta^\mu_\sigma\,\delta^\vrho_\nu$. 

A vector is said to be \emph{covariant} if its infinitesimal transformation under a generalised diffeomorphism generated by a vector with components $\xi^M$ is given by the generalised Lie derivative generated by $\xi^M$ 
\begin{equation}\label{deltaXi}
\delta_\xi\,V^M = L_{\xi,\sigma(\xi)}\,V^M,
\end{equation}
for some ancillary parameter $\sigma_M(\xi)$ written in terms of $\xi^M$. Define $\Delta_\xi$ as the breaking of the covariance 
\begin{equation}
\Delta_\xi := \delta_{\xi}-L_{\xi,\sigma(\xi)}.
\end{equation}

The definition of covariance in \eqref{deltaXi} requires that the Jacobian matrix of a change of generalised coordinates $Y^M$ is\footnote{Consistently, $M^M{}_N \rightarrow \de_\nu\,\xi^\mu$ in ordinary geometry.}
\begin{equation}
\tfrac{\de Y^M}{\de Y'^N} = \delta^M_N + Z^{MP}{}_{QN}\,\de_P\,\xi^Q + f^{MP}{}_N\,\sigma_P(\xi) =: \delta^M_N - M^M{}_N.
\end{equation}
In this way, the transformation of the partial derivative of a covariant vector $V^M$ is
\begin{align}
\de'_M\,V'^N(Y) &= \tfrac{\de Y^R}{\de Y'^M}\,\de_R\,(\tfrac{\de Y^N}{\de Y'^Q}\,V^Q(Y')) = \nn\\
&= \tfrac{\de Y'^R}{\de Y^M}\,\tfrac{\de Y^N}{\de Y'^Q}\,\de_R\,V^Q(Y') + \tfrac{\de Y'^R}{\de Y^M}\,\de_R\,\tfrac{\de Y^N}{\de Y'^Q}\,V^Q(Y'),
\end{align}
where the second term parametrises the breaking of the covariance of the partial derivative of a covariant vector. Correspondingly, the infinitesimal transformation reads
\begin{equation}
\delta_{\xi,\sigma(\xi)}\,V^M(Y) = 
L_{\xi,\sigma(Y)}\,\de_M\,V^N(Y) + \delta^R_M\,\de_R\,(\delta^N_Q - M^N{}_Q)\,V^Q(Y),
\end{equation}
which gives the breaking of the covariance 
\begin{equation}\label{SpoilingPartialV}
\Delta_\xi\,\de_M\,V^N(Y) = -\de_M\,M^N{}_Q\,V^Q(Y).
\end{equation}
If a covariant derivative $\nabla_M$ with respect to a connection $\Gamma_M{}^N{}_P$, 
\begin{equation}
\nabla_M\,V^N = \de_M\,V^N + \Gamma_M{}^N{}_P\,V^P
\end{equation}
is introduced, requiring that the covariant derivative of a covariant vector is a covariant vector too
\begin{equation}
\Delta_\xi\,\nabla_M\,V^N = 0,\;\;\text{if}\;\;\Delta_\xi\,V^N = 0,
\end{equation}
then the breaking of the covariance of the connection has to be equal to \cite{Cederwall:2015ica}
\begin{equation}\label{DeltaXiGamma}
\Delta_\xi\,\Gamma_M{}^N{}_P = 
\de_M\,M^N{}_P = -Z^{NR}{}_{QP}\,\de_M\,\de_R\,\xi^Q - f^{NR}{}_P\,\de_M\,\sigma_R(\xi)
\end{equation}
in order to compensate \eqref{SpoilingPartialV}.

In $n=8$ exceptional geometry, $\Gamma_M{}^N{}_P$ is contained in $\bm{248} \otimes (\bm{1} \oplus \bm{248})$, where $\bm{1}$ is the $\mathbb{R}^+$ component in the duality group. The decomposition in irreducible components reads \cite{Cederwall:2015ica}
\begin{equation}
\bm{248} \otimes (\bm{1} \oplus \bm{248}) = 
\bm{248} \oplus (\bm{1} \oplus \bm{3875} \oplus \bm{27000}) \oplus (\bm{248} \oplus \bm{30380}),
\end{equation}
where the sum $\bm{1} \oplus \bm{3875} \oplus \bm{27000}$ is the symmetric part of $\bm{248} \otimes \bm{248}$, and $\bm{248} \oplus \bm{30380}$ is the antisymmetric part. On the other hand, $\Delta_\xi\,\Gamma_M{}^N{}_P$ is also contained in $\bm{248} \otimes \bm{27000}$, since its leading term is $\de_M\,\de_P\,\xi^N$, whose lowered indices are on section; therefore, only the $\bm{27000}$ component in the symmetrised product of $\bm{248} \otimes \bm{248}$ survives, since the section constraint sets  $\bm{1}\oplus\bm{3875}$ to zero. Because of the following decomposition
\begin{equation}
\bm{248} \otimes \bm{27000} = 
\bm{248} \oplus \bm{27000} \oplus \bm{30380} \oplus \bm{779247} \oplus \bm{1763125} \oplus \bm{4096000},
\end{equation}
the common part with $\bm{248} \otimes (\bm{1} \oplus \bm{248})$ is $\bm{248} \oplus \bm{27000} \oplus \bm{30380}$ \cite{Cederwall:2015ica}. Now, the torsion $T_M{}^N{}_P(\Gamma)$, defined by \cite{Coimbra:2011ky}
\begin{equation}
({L_{\xi,\sigma(\xi)}}^{(\de_M \rightarrow \nabla_M)} - L_{\xi,\sigma(\xi)})\,V^N = 
-\xi^M\,V^P\,T_M{}^N{}_P(\Gamma), 
\end{equation} 
where the first piece is the Lie derivative with the partial derivatives formally replaced by the covariant ones, is the covariant part of the connection $\Gamma_M{}^N{}_P$:
\begin{equation}
\Delta_\xi\,T_M{}^N{}_P(\Gamma) = 0.
\end{equation}
From the previous analysis, one concludes that the torsion is contained in $\bm{1} \oplus \bm{248} \oplus \bm{3875}$, which are the components of the embedding tensor in three-dimensional gauged supergravity as expected in $E_8$ exceptional field theory \cite{Hohm:2014fxa}.

If $\Gamma_M{}^N{}_P$ is chosen to be the Weitzenb\"ock connection $W_M{}^N{}_P = E^N{}_A\,\de_M\,E_P{}^A$ for any choice of the frame $E^M{}_A$, then the torsion is the intrinsic torsion \cite{Inverso:2024xok}
\begin{equation}
T_M{}^N{}_P = T_M{}^N{}_P(W) = - E_M{}^A\,E_P{}^B\,L_{E_A,h_A}\,E^N{}_B,
\end{equation}
where one sets $\xi^M = E^M{}_A\,\tilde{\xi}^A$, $V^M = E^M{}_A\,\tilde{V}^A$, for some $\tilde{\xi}^A$, $\tilde{V}^A$, and the ancillary parameter $h_A(\xi)$ parametrises $\sigma(\xi)$, according to
\begin{equation}
E_M{}^A\,h_{NA}\,\xi^M = \sigma_N(\xi).
\end{equation}
One can show that $h_A(\xi)$ is fixed by the condition that $T_M{}^N{}_P$ sits in $\bm{1} \oplus \bm{248} \oplus \bm{3875}$. The result is \cite{Inverso:2024xok, Rovere:2025jmp}
\begin{equation}\label{FixedSigma}
\sigma_N(\xi) = E_M{}^A\,h_{NA}\,\xi^M = \tfrac{1}{60}\,\xi^M\,f_{MP}{}^Q\,W_N{}^P{}_Q.
\end{equation}
According to such expression, the index of $\sigma_M(\xi)$ is consistently on section.  

Using \eqref{DeltaXiGamma} with $\Gamma_M{}^N{}_P \rightarrow W_M{}^N{}_P$, \eqref{ZE8}, \eqref{FixedSigma}, and \eqref{DecompositionW}, one gets the breaking of the covariance of the components of the Weitzenb\"ock connection:
\begin{equations}
\Delta_\xi\,W'_M &= -2\,\de_M\,\de_N\,\xi^N,\label{SpoilW1}\\
\Delta_\xi\,W''_{MN} &= -\de_M\,(f_N{}^P{}_Q\,\de_P\,\xi^Q + W''_{NP}\,\xi^P).
\end{equations}
As a check, one can compute the breaking of the covariance of the components of the torsion, using \eqref{theta0}--\eqref{theta2}:
\begin{equations}
\Delta_\xi\,\vtheta &= -\tfrac{1}{4}\,\eta^{MN}\,\de_M\,(W''_{NP}\,\xi^P),\\
\Delta_\xi\,\vtheta'_M &= (\tfrac{1}{2}\,f_{MN}{}^{PR}\,\de_P\,\de_R-2\,\de_M\,\de_N)\,\xi^N - f^{PQ}{}_M\,\de_P\,(W''_{QN}\,\xi^N),\\
\Delta_\xi\,(\vtheta''_{MN}+\eta_{MN}\,\vtheta) &= (\tfrac{1}{2}\,f_{MN}{}^{PQ}-\delta^P_M\,\delta^Q_N - \delta^P_N\,\delta^Q_M)\,\de_P\,(W''_{QR}\,\xi^R)\nn\\
& -(\tfrac{1}{2}\,f_{MN}{}^{PQ}\,f_{S}{}^{R}{}_Q\,\de_R\,\de_P-f_{MS}{}^{R}\,\de_N\,\de_R - f_{NS}{}^{R}\,\de_M\,\de_R)\,\xi^S.
\end{equations}
As expected, they vanish on section (one has to use \eqref{FsymSecCon} for the last one):
\begin{equation}
\Delta_\xi\,\vtheta \approx 0, \quad
\Delta_\xi\,\vtheta'_M \approx 0, \quad
\Delta_\xi\,\vtheta''_{MN} \approx 0.
\end{equation}

\section{Derivation of the Bianchi identities}\label{App3}

Define the following tensor:
\begin{align}
B_{MN}{}^P{}_Q &:= (D^{(-1/2)}_M + \delta_{T_M})\,T_N{}^P{}_Q - T_N{}^P{}_Q \big{|}_{W_N{}^P{}_Q \rightarrow D^{(-1/2)}_N\,T_M{}^P{}_Q} \nn\\
&= D^{(-1/2)}_M\,T_N{}^P{}_Q 
+ T_M{}^R{}_Q\,T_N{}^P{}_R 
- T_M{}^R{}_Q\,T_M{}^P{}_R \nn\\
& + T_M{}^R{}_M\,T_R{}^P{}_Q 
- D^{(-1/2)}_N\,T_M{}^P{}_Q 
+ D^{(-1/2)}_Q\,T_M{}^P{}_N  \nn\\
& - Y^{PM}{}_{SQ}\,D^{(-1/2)}_R\,T_M{}^S{}_N - \tfrac{1}{60}\,f^{RP}{}_Q\,f_{NS}{}^T\,D^{(-1/2)}_R\,T_M{}^S{}_T,
\end{align} 
which vanishes using the Bianchi identity \eqref{BianchiT}. Tracing the first and the second pairs of indices, one finds the identity \eqref{Bianchi4}:
\begin{equation}
-\tfrac{1}{31}\tfrac{1}{248}\,B^M{}_M{}^P{}_P = 
D^{(-1/2)}_M\,\vtheta'^M - \vtheta'_M\,\vtheta'^M.
\end{equation}
Tracing the first pair of indices and dualising the last one with the structure constants, one finds the identity \eqref{Bianchi1}:
\begin{equation}
\tfrac{8}{60}\tfrac{1}{248}\,f_{LP}{}^Q\,B^M{}_M{}^P{}_Q = 
f^{MN}{}_L\,D^{(-1/2)}_M\,\vtheta'_N + 8\,D^{(-1/2)}_L\,\vtheta - \vtheta'^M\,\vtheta''_{ML} - \vtheta\,\vtheta'_L =: b'_L.
\end{equation}
The identity \eqref{Bianchi2} is obtained by dualising the first pair of indices and tracing the last one, and subtracting the previous identity:
\begin{equation}
\tfrac{5}{60}\tfrac{1}{248}\,f^{MN}{}_L\,B_{MN}{}^P{}_P -  b'_L =
D^{(-1/2)}_M\,\vtheta''^M{}_L -7\,D^{(-1/2)}_L\,\vtheta + \tfrac{1}{2}\,\vtheta'^M\,\vtheta''_{ML} - \tfrac{7}{2}\,\vtheta\,\vtheta'_L =: b''_L.
\end{equation}
Finally, antisymmetrising the first pair of indices, tracing the last two, and subtracting the dual $b'_L$, one gets the last identity \eqref{Bianchi3}:
\begin{align}
-\tfrac{1}{248}\,B_{[MN]}{}^P{}_P - \tfrac{1}{2}\,f_{MN}{}^L\,b'_L =&\,
(D^{(-1/2)}_R-\vtheta'_R)\,\vtheta''_{L[M}\,f_{N]}{}^{LR} + 3\,D^{(-1/2)}_{[M}\,\vtheta'_{N]} \nn\\
& + 2\,f_{MN}{}^L\,D^{(-1/2)}_L\,\vtheta +\tfrac{3}{2}\,f_{MN}{}^L\,\vtheta\,\vtheta'_L \nn\\
& - \tfrac{1}{2}\,f_{MN}{}^R\,\vtheta'^L\,\vtheta''_{LR} =: b''_{MN}.
\end{align}
One can observe that $b'_L$, $b''_L$, and $b'''_{MN}$ are not independent, since the following identity holds:
\begin{equation}
b'''_{MN}\,f^{MN}{}_L = 6\,(b'_L-4\,b''_L).
\end{equation}
$B_{MN}{}^P{}_Q$ also has a symmetric part in the first pair of indices, which, similarly to the antisymmetric part considered before, gives $f_P{}^Q{}_{(M}\,D^P\,\vtheta''_{N)P}$ in terms of $D_{(M}\,\vtheta_{N)}-\tfrac{1}{2}\,f_{MN}{}^{PQ}\,D_P\,\vtheta'_Q$. Combining the symmetric and the antisymmetric parts, one gets a Bianchi identity that fixes $f^{PQ}{}_M\,D_P\,\vtheta''_{NQ}$ in terms of $D_L\,\vtheta$ and $D_M\,\vtheta'_N$:
\begin{align}
f^{PQ}{}_M\,D_P\,\vtheta''_{NQ} =&\, 
-2\,D_M\,\vtheta'_N + D_N\,\vtheta'_M 
+\tfrac{1}{4}\,f_{MN}{}^{PQ}\,D_P\,\vtheta'_Q
-f_{MN}{}^P\,D_P\,\vtheta \nn\\
& +\vtheta'_M\,\vtheta'_N 
- \tfrac{1}{4}\,f_{MN}{}^{PQ}\,\vtheta'_P\,\vtheta'_Q + \tfrac{1}{4}\,f_{MN}{}^P\,\vtheta'^P\,\vtheta''_{QP} \nn\\
& -\tfrac{3}{4}\,f_{MN}{}^P\,\vtheta\,\vtheta'_P + f^{PQ}{}_M\,\vtheta'_P\,\vtheta''_{NQ}.
\end{align}

\bibliographystyle{JHEP}
\bibliography{ir_7-26}

\end{document}